\documentclass[journal,comsoc]{IEEEtran}

\usepackage[T1]{fontenc}

\usepackage{flushend}

\usepackage{cite}

\usepackage{textcmds}

\usepackage{blindtext}

\usepackage{xcolor}
\usepackage{hyperref}
\usepackage[all]{hypcap}
\hypersetup{colorlinks=true, pdfstartview=FitV, linkcolor=blue, citecolor=blue, plainpages=false, urlcolor=blue}

\usepackage{multirow}

\ifCLASSINFOpdf
   \usepackage[pdftex]{graphicx}
\else
\fi

\usepackage{amsmath}

\usepackage[cmintegrals]{newtxmath}

\begin{document}

\title{Dependable Intrusion Detection System for IoT: A Deep Transfer Learning-based Approach}
\author{Sk. Tanzir Mehedi, \IEEEmembership{Member, IEEE}, Adnan Anwar, \IEEEmembership{Member, IEEE}, Ziaur Rahman, \IEEEmembership{Member, IEEE}, Kawsar Ahmed, \IEEEmembership{Member, IEEE}, and Rafiqul Islam, \IEEEmembership{Senior Member, IEEE}
\thanks{Manuscript received October 16, 2021; revised March 31, 2022.}
\thanks{Sk. T. Mehedi is with the Department of Information and Communication Technology (ICT), Mawlana Bhashani Science and Technology University (MBSTU), Tangail 1902, Bangladesh (e-mail: tanzirmehedi@ieee.org). }
\thanks{A. Anwar is with the Centre for Cyber Security Research and Innovation (CSRI), Deakin University, Geelong 3216, Australia (e-mail: adnan.anwar@deakin.edu.au).}
\thanks{Z. Rahman is with the School of Computing Technologies, RMIT University, Melbourne 3001, Australia (e-mail: rahman.ziaur@rmit.edu.au).}
\thanks{K. Ahmed is with the Department of ICT, MBSTU, Tangail-1902, Bangladesh and Department of ECE, University of Saskatchewan, 57 Campus Drive, Saskatoon, SK S7N 5A9, Canada (e-mail: kawsar.ict@mbstu.ac.bd).}
\thanks{R. Islam is with the School of Computing and Mathematics, Charles Sturt University, Albury, NSW 2640, Australia (e-mail: mislam@csu.edu.au).}}

\maketitle

\begin{abstract}

Security concerns for IoT applications have been alarming because of their widespread use in different enterprise systems. The potential threats to these applications are constantly emerging and changing, and therefore, sophisticated and dependable defense solutions are necessary against such threats. With the rapid development of IoT networks and evolving threat types, the traditional machine learning-based IDS must update to cope with the security requirements of the current sustainable IoT environment. In recent years, deep learning, and deep transfer learning have progressed and experienced great success in different fields and have emerged as a potential solution for dependable network intrusion detection. However, new and emerging challenges have arisen related to the accuracy, efficiency, scalability, and dependability of the traditional IDS in a heterogeneous IoT setup. This manuscript proposes a deep transfer learning-based dependable IDS model that outperforms several existing approaches. The unique contributions include effective attribute selection, which is best suited to identify normal and attack scenarios for a small amount of labeled data, designing a dependable deep transfer learning-based ResNet model, and evaluating considering real-world data. To this end, a comprehensive experimental performance evaluation has been conducted. Extensive analysis and performance evaluation show that the proposed model is robust, more efficient, and has demonstrated better performance, ensuring dependability.

\end{abstract}


%
\IEEEpeerreviewmaketitle

\section{Introduction}
\label{sec:introduction}
\IEEEPARstart{S}{ecurity} is an essential component of a dependable system. Since the inception of the Industry 4.0 revolution early last decade, the integration of intelligent Internet of Things (IoT) sensors has become inevitable to keep running the latest enterprise ecosystem smoothly. Thus, the popularity of IoT devices has greatly increased over the last couple of years because of their appealing connectivity, communication, and decision-making features \cite{Eirini-1}. Admittedly, the rise of cyber-attacks has increased in line with the massive online deployments of IoT devices \cite{Da-9}. However, the cyber-physical world is not always predictable \cite{Redowan-73}. Although the interactions within integrated cyber-physical systems are complex, there is no doubt that they must be "dependable" \cite{Bhuiyan-4}. Future heterogeneous IoT applications must close the dependability gap in the face of various challenges and services, such as the ability to overcome data anomalies, avoid catastrophic failures, avoid data corruption, and prevent deliberate privacy intrusions, among others \cite{Bhuiyan-4, 9594795}. As industrial heterogeneous IoT devices have proliferated (25.1 billion by 2025 forecast by the GSM Association). This, has had a direct impact on our socio-economic lives \cite{Alsaedi-39, Manos-6}. So, security and privacy issues relating to industrial heterogeneous IoT devices need to be addressed by ensuring dependability with care and attention.

Generally, IoT-based applications consist of lightweight communication protocols with limited computational power and storage capacity \cite{Emiliano-10, hasan-75}. However, traditional security mechanisms require high computational capabilities \cite{Da-9}. As a result of their heterogeneous nature and characteristics, this mechanism cannot be directly deployed for IoT-based applications \cite{Da-9, Emiliano-10}. As IoT applications operate inside the network, this mechanisms seem to be insufficient because of their characteristics, which only identify external attacks \cite{Tianlong-11}. Because of the overwhelming rise of IoT threats and vulnerabilities, they have already shown their inadequate strength to fight against targeted attacks and associated anomalies \cite{Jayavardhana-13}. As a result, the potential threats to the IoT ecosystem are constantly emerging and changing, and the exact nature of these threats or attacks is complex to investigate.

Furthermore, traditional machine learning (TML) algorithms have advanced significantly, and some of their variants have been successfully applied to solve classification tasks related to intrusion detection \cite{Nisioti-67, Alsaedi-74}. Also, the progression of deep learning (DL) and its great success in different fields has guided it as a potential solution for network intrusion detection \cite{Nisioti-67}. However, most of proposed IDSs focus on detecting a limited number of cyber-attack types and device profiling and do not consider automatic attack type classification and dependability \cite{Li-32}. Most importantly, these models require a large amount of training data, and the size of training datasets affects the accuracy of the model when the training data is insufficient, as highlighted by the authors in \cite{Anca-8, Shahid-3}. As a result, the TML and DL models need sufficient training data, and it is difficult to train a dependable IDS model only depending on a small-scale of target domain data.

\begin{figure}[htp]
    \centering
    \includegraphics[width=\columnwidth]{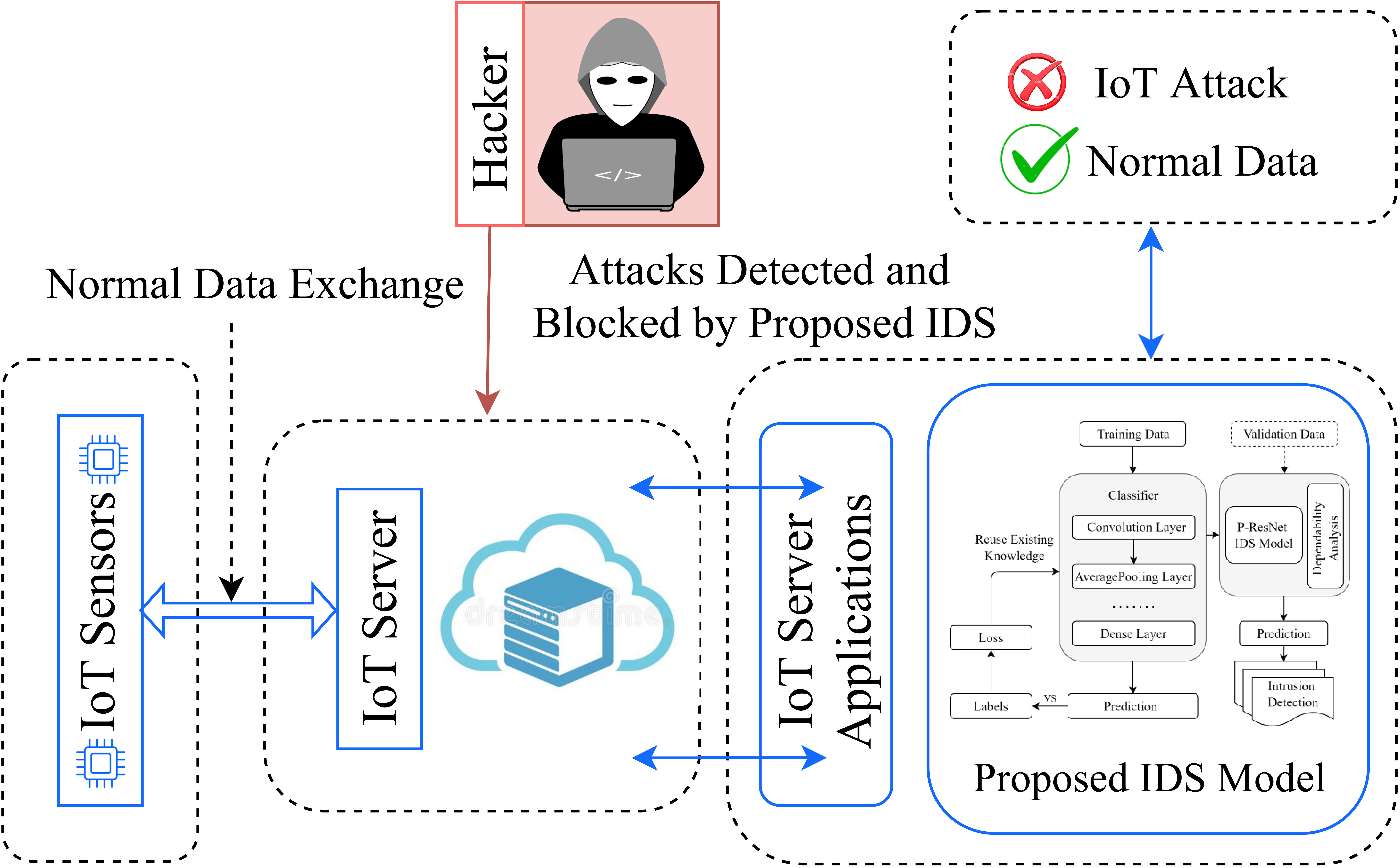}
    \caption{The overall scenario of the proposed IDS applied to the heterogeneous IoT network}
    \label{The overall scenario of the proposed IDS applied to the heterogeneous IoT network}
\end{figure}

With the rapid development of IoT networks and the emergence of new threat types, deep transfer learning (DTL) models, which have effectively solved many nonlinear problems, can be used as a promising solution for building a dependable IDS for the IoT ecosystem \cite{Jun-18}. Specifically, in the Industry 4.0 revolution, heterogeneous IoT applications require a high level of dependable security structure \cite{Bhuiyan-17}. Therefore, there is a need for a dependable intrusion detection system to identify malicious activity within any heterogeneous IoT network. Dependability performance analysis includes availability, efficiency, and scalability features \cite{Bhuiyan-4}. An IDS model's availability is determined by how often it is available, and its efficiency is how well it performs with low computational complexity \cite{Bhuiyan-4}. Also, scalability is the capability to adapt easily to the increased number of data sources from a wider range of IoT sensors \cite{Gregory-4}. To consider all the dependencies, we propose a deep transfer learning-based residual neural network (P-ResNet) IDS that can train efficiently using only a small scale of target domain data and understand its impact by ensuring dependability with low computational capabilities. Figure~\ref{The overall scenario of the proposed IDS applied to the heterogeneous IoT network} shows the motivation on how the proposed IDS can be applied to the IoT network toward effective attack detection.

The proposed method can accurately classify between normal and attack scenarios without any type of failure or undergoing a repair action, which maintains the availability. Moreover, the proposed model increased the scalability properties by including various heterogeneous trusted data sources in the training dataset with maximum consistency that were acquired from a wider range of IoT sensors. Finally, our proposed model improves performance compared to several other existing approaches. The overall accuracy of the proposed detection model is 87\% with low computational complexity. Moreover, the model also greatly improves the precision score of 88\%, the recall score of 86\%, and the f1-score of 86\%, which are higher than the benchmark models. This demonstrates that the proposed detection method is more efficient than others. Extensive analysis and performance evaluation show that the proposed model is robust, more efficient, and has demonstrated better performance, ensuring dependability. The key contributions of this paper are summarized as follows:

\begin{itemize}

\item 
In this work, a deep transfer learning-based residual neural network (ResNet) has been proposed for effective intrusion detection. This model effectively detect various cyber threats against heterogeneous IoT networks. Such threats include, denial-of-service (DoS), distributed denial-of-service (DDoS), data injection, man-in-the-middle (MITM), backdoor, password cracking attack (PCA), scanning, cross-site scripting (XSS), and ransomware attack.

\item 
The experiments have been conducted using IoT data generated from heterogeneous sources that include seven IoT sensors. We have observed empirical IoT stream data to identify the best features in the context of supervised learning for speeding up training and improving performance metrics while ensuring its dependability.

\item Finally, comprehensive experimental performance analyses have been conducted considering various deep learning and deep transfer learning algorithms. Extensive analysis and performance evaluation show that the proposed model is more dependable and outperforms others. 

\end{itemize}

The rest of this research paper is organized as follows. First, Section \ref{sec: related work} discusses the background and related works. In Section \ref{sec: proposed methodology}, we present the details of our proposed methodology. Section \ref{sec: model implementation and evaluation} presents a detailed description of the dataset and the training process of the selected models. Furthermore, evaluation and experimental results analysis of the methods are discussed in Section \ref{sec: result analysis}. Finally, the summary and possible future research directions are given in Section \ref{sec: conclusion and future work}.

\section{Related Work}
\label{sec: related work}

While investigating the most recent relevant works, we found several works that overlap similar motivations differently. The following discussion illustrates those works before we demonstrate our proposed methodology.

Deep transfer learning is a solution that can reuse existing knowledge from the previously trained model and achieve better intrusion detection performance than other models \cite{Li-32}. The transfer learning model proposed by Zadrozny et al. \cite{Zadrozny-34} for intrusion detection, shows more efficient performance in both labeled and unlabeled data. Dai et al. \cite{Dai-35} also proposed a transfer learning model called TrAdaBoost. This model allows knowledge to be efficiently transferred from the old trained data to construct a more efficient classification model for the new validation data. Raina et al. \cite{Raina-36} proposed a transfer learning model that builds an informative model on existing knowledge before validating a particular new task. Gou et al. \cite{Gou-37} proposed a novel transfer learning algorithm to detect different cyber-attacks on large-scale IoT ecosystems. The experimental analysis shows that the detection accuracy of the proposed model for different types of cyber-attacks has been comprehensively improved. 

Furthermore, Li et al. \cite{Li-32} proposed a transfer learning approach for intrusion detection of different types of attacks on the Internet of Vehicles (IoV). The experimental analysis shows that the model has greatly improved the detection accuracy rate by 23\%. Xu et al. \cite{Xu-38} proposed an intrusion detection approach based on deep learning and transfer learning. Here, transfer learning is initiated to enhance the efficiency and adaptability of the model. The experimental analysis shows that the proposed model is more efficient and robust than other methods and performs better in detecting and classifying new cyber-attacks more effectively. In addition, a real-time, dependable method for human activity classification based on transfer learning has been proposed by Wan et al. \cite{Wan-32}. The experimental analysis shows that the proposed method outperforms others. Recently, Jiang et al. \cite{Jiang-33} proposed a dependable deep learning-based model design for IoT load identification in smart grid. The experimental results demonstrate the good generalization performance and superiority for NILMI in IoT load monitoring systems. However, further analyses are necessary to investigate the performance of new complex types of cyber-attacks in various categories. A summary of all the existing literature that overlap similar motivations differently is given in Table \ref{Table: Overview of recent research on IDSs for IoT applications}. The literature is organized according to \textit{detection\_category}, \textit{model\_accuracy}, \textit{proposed\_methodology}, \textit{validation\_strategy}, and \textit{dependability\_performance\_analysis} of the proposed model. 

To get the optimal performance of existing transfer-learning models and maintain the knowledge transfer process in realistic models for heterogeneous IoT ecosystems, the deep computational intelligence system has recently been applied in transfer-learning \cite{Lu-33}. So, the existing transfer learning solutions for intrusion detection still require a small amount of labeled data for model updates in heterogeneous IoT applications, ensuring dependability. A new-generation labeled \textit{TON\_IoT\_Telemetry\_Dataset} of IoT devices for data-driven IDS was proposed by Alsaedi et al. \cite{Alsaedi-39}, which is more suitable for applying deep transfer learning models. The deep transfer learning approach shows better performances in time series classification than other models \cite{Xu-38}. Because of the original situation of heterogeneous IoT applications, this paper has comprehensively improved the existing transfer learning model to ensure its dependability for detecting various complex types of cyber-attacks.

\begin{table*}[htp]
\caption{Overview of recent research on IDSs for IoT applications. "C" indicates that the topic is covered, "N" indicates that the topic is not covered and "P" indicates that the topic is partially covered.}
\label{Table: Overview of recent research on IDSs for IoT applications}
\centering
\begin{tabular}{|c|c|c|c|c|c|}
\hline
Ref. & Detection Category & Accuracy & Proposed Methodology & Validation Strategy & Dependability\\
\hline
\cite{Zadrozny-34} & Learning and Evaluating Classifiers & $>$90\% & Transfer Learning & Analytical and Experimental & N\\
\hline
\cite{Dai-35}      &  Efficiently Transferred Trained-data  & $>$85\% & Transfer Learning & Theoretically and Empirically & N\\
\hline
\cite{Raina-36}    & Intrusion Detection System  & $>$85\% & Transfer Learning & Simulation & N\\
\hline
\cite{Gou-37}      & Intrusion Detection System  & $>$90\% & Distributed Transfer Learning & Empirical (6 Devices) 4 Categories & P\\
\hline
\cite{Li-32}       & Intrusion Detection System  & $>$85\% & Transfer Learning & Empirical & N\\
\hline
\cite{Xu-38}       & Intrusion Detection System & $>$80\% & Deep Transfer Learning & Empirical (4 Devices) 6 Categories & P\\
\hline
\cite{Jiang-33}    & Device Load Monitoring System & $>$85\% & Transfer Learning & Experimental & N\\
\hline
\cite{Bhuiyan-17}  & Decision-Making in CPSs  & $>$85\% & Model-based In-network & Experimental & C\\
\hline
Proposed           & Intrusion Detection System &  $>$87\%& Deep Transfer Learning & Empirical (7 Devices) 9 Categories & C\\
\hline
\end{tabular}
\end{table*}

\section{Proposed Methodology}
\label{sec: proposed methodology}

This section presents our proposed dependable IDS model of a deep transfer learning-based residual neural network (P-ResNet). We first introduce the problem statement. Next, we thoroughly explain the structure of the proposed model and how we adapted the model for the DTL process.

\subsection{Problem Statement}

Six key motivations studied inspired this research paper. First, most proposed IDSs focus on detecting a limited number of cyber-attacks \cite{Eirini-1}. In this case, our proposed model's target is to identify and classify a wide range of cyber-attacks efficiently (e.g., DoS, data injection, MITM, backdoor, PCA, scanning, and ransomware). Such attacks are significantly more challenging to detect at low computational complexity.

There is limited device profiling in the existing research\cite{Li-32}. Therefore, this paper considers the behavior of seven different IoT devices (e.g., fridge, GPS tracker, motion light, garage door, modbus, thermostat, and weather sensor) to increase the integrity properties of our proposed DTL-based P-ResNet model. 

Furthermore, most proposed IDSs did not consider automatic attack type classification and dependability \cite{Alsaedi-39}. However, our proposed DTL-based P-ResNet model can automatically detect different attacks and efficiently classify different attack scenarios while ensuring dependability.

Most importantly, the existing TML and DL-based IDSs have apparent advantages in detecting various cyber-attacks \cite{Wang-22}. However, these models require a large amount of training data, and the size of training datasets affects the accuracy of the model when the training data is insufficient, as has been confirmed by the subsequent experiment \cite{Anca-8}. As a result, the TML and DL models need sufficient training data, and it is challenging to train an efficient IDS model only depending on a small-scale of target domain data.

Moreover, IoT-based applications typically consist of lightweight communication protocols with limited computational power and storage capacity \cite{Emiliano-10}. However, traditional security mechanisms require high computational capabilities \cite{Anca-8}. As a result, these conventional security mechanisms cannot be directly deployed for IoT-based applications due to their heterogeneous nature and characteristics (e.g., less scalability, inefficiency) \cite{Da-9, Emiliano-10}. Furthermore, as IoT applications operate inside the network, traditional security mechanisms seem to be insufficient because of their features, which only identify external attacks \cite{Tianlong-11}. Because of the overwhelming rise of IoT threats and vulnerabilities, traditional security mechanisms have already shown their inadequate strength to fight against targeted attacks, and associated anomalies \cite{Jayavardhana-13}. As a result, the potential threats to the heterogeneous IoT ecosystem are constantly emerging and changing, and the exact nature of these threats is unknown.

To address the last two dependencies of the TML and DL models, we propose a dependable deep transfer learning-based residual neural network (P-ResNet) IDS that can train efficiently using only a small scale of target domain data and understand its impact with low computational capabilities. The proposed P-ResNet model significantly enhances the performance of neural networks with more layers. Generally, our proposed model tries to tackle those dependencies by directly bypassing the input data to the output layer, allowing the output layer to access the raw data. Therefore, we transfer the knowledge in source domain data to the target domain through the proposed P-ResNet model and combine the target domain data with the same model to construct an efficient and dependable IDS for the target domain to improve the detection accuracy for any heterogeneous IoT ecosystem.

Finally, most of the previously proposed IDSs didn't consider dependability performance analysis \cite{Bhuiyan-4}. In this case, we have analyzed the dependability performance to ensure the availability, efficiency, and scalability features of our proposed deep transfer learning-based P-ResNet IDS model.

\subsection{ResNet with Deep Transfer Learning}

Deep neural network (DNN) depth is an essential aspect of model performance\cite{Li-32, Lu-33}. More complex features may be extracted by a DNN algorithm with more layers \cite{Xu-38}. As a result, deeper DNNs may theoretically result in better performance. If the number of DNN layers increases, the degradation problem causes the DNN algorithms' to become more saturated \cite{Alsaedi-39}. The deep residual network (ResNet) tries to tackle this problem by directly bypassing the input data to the output layer, allowing the output layer to access the raw data. Our proposed P-ResNet model consists of multiple residual blocks, each of which may be defined as- $U(x) = V(x) + x$, where x denotes the input data, V(x) denotes the identity residuals mapping function, and U(x) denotes the mapped solution function.

Deep transfer learning is a supervised machine learning technique that uses a small-scale dataset to apply a pre-trained model \cite{Xu-38}. The definition of deep transfer learning has been given in terms of source and target domains. A labeled domain IoT dataset $D$ consists of data instances $X$ and a label $Y$. A task $T$ is defined as the connection of label $Y$ with a prediction function $f$ to be learned from the labeled domain data. The source domain and target domain datasets are denoted by $D_s = (X_s, Y_s)$ and $D_t = (X_t, Y_t)$, while the source tasks and target tasks are denoted by $T_s = (Y_s, f_s)$ and $T_t = (Y_t, f_t),$ respectively. We assume that $T_t$ and $D_t$, which represent the ubiquitous computing issue to be solved and the corresponding labeled dataset, are accessible.

\textbf{Source Domain:} The domain where the initial model is located. The source domain data ($D_s : (X_s,Y_s)$) is the combination of $ {(X_{s1},Y_{s1}), (X_{s2},Y_{s2}),...,(X_{sn}, Y_{sm})}$, in which the class of the source domain ($Y_s$) is {0, and 1}, where the normal and attack scenarios are represented by 1 and 0, respectively.

\textbf{Target Domain:} The domain has a new type of attack. The target domain data ($D_t : (X_t,Y_t)$) is the combination of $ {(X_{t1},Y_{t1}), (X_{t2},Y_{t2}),...,(X_{tn},Y_{tm})}$, in which the class of target domain ($Y_t$) is {0 and 1}, where the normal scenario is represented by 1 and the attack scenario is represented by 0.

The number of source and target domain data is represented by n and m, respectively. Furthermore, the source domain label ($ Y_s $) and the target domain label ($ Y_t $) contain only ``normal” and ``attack” data, but attackers in the source domain and target domain may be different. Although the source domain label ($ Y_s $) and the target domain label ($ Y_t $)  have the same feature space, their performance in specific features is different. To measure the difference between the source domain and the target domain, we have used the following formula~\eqref{eq1}, which is called Maximum Mean Discrepancy (MMD) \cite{Nisioti-67}.

\setlength{\abovedisplayskip}{0pt}
\setlength{\belowdisplayskip}{6pt}

\begin{align}
\label{eq1}
Dist (X_s, X_t) = \Bigg\| \frac{1}{n} \sum_{i=0}^{n} \phi (X_{s_i}) - \frac{1}{m} \sum_{i=0}^{m} \phi (X_{t_i}) \Bigg\|^2 
\end{align}

We propose a deep transfer learning method that transfers DNN weights learned on different attack classification problems on $D_s$ to another DNN trained to solve $T_t$ on $D_t$. Our proposed method in Figure \ref{The structure of the proposed P-ResNet model} belongs to the category of inductive transfers since the source tasks and target tasks are distinct $(T_s \neq T_t)$. The steps are as follows:

\begin{itemize}
    \item \textbf{Definition of source domain $\mathbf{D_s}$ and source task $\mathbf{T_s}$:} $X_s$ is firstly built by considering $M_c$ multiple channel IoT datasets (e.g. $modbus, motion\_light, weather$ sensor etc.). Each  $M_c$  sequence in the $n^{th}$ dataset $(1 \leq n \leq M_c)$ is decomposed into single channels, each of which is divided into different segments $S$ of length $L$ using a \textit{sliding\_window\_approach}. All the segments are combined to form the source dataset $X_s$, which is defined by the following formula~\eqref{eq2}: 
    
    \begin{align}
    \label{eq2}
    X_s = \bigcup_{n=1}^{M_c}\left \{ x_{i}^{(n)} \epsilon  \mathbb{R^{L}} | \left (  1 \leq  i \leq  S_{n}\right ) \right \}
    \end{align}
     
     Here, $x_{i}^{(n)}$ refers to the $i^{th}$ segment of the $n^{th}$ source dataset $X_s$, and $S_n$ refers to the total number of segments obtained from the $n^{th}$ source dataset $X_s$. That is to say, $X_s$ is the union of all segments $S$ extracted from the multiple channel $M_c$ source datasets $X_s$. The source task $T_s$ is defined as the classification of different attacks on the source domain $D_s$. Source labels $Y_s$ are defined by the following formula~\eqref{eq3}:
    
    \begin{align}
    \label{eq3}
        Y_s = \bigcup_{j=1}^{M_c}\left \{ y_{i}^{(n)} \epsilon \left \{ 1, 2,., C_s \right \} | \left (  1 \leq  i \leq  S_{n}\right ) \right \}
    \end{align}
     Here, $C_s$ is the number of classes of the source domain (e.g., normal and attack), and $y_{i}^{(n)}$ indicates the label of $x_{i}^{(n)} \epsilon X_s$. Furthermore, $f_s$ is the source prediction function which attributes each $x_{i}^{(n)}$ to its corresponding label $y_{i}^{(n)}$, whether normal or attack.
     \vspace{2.5pt}
     \item \textbf{Learning of source prediction  function $\mathbf{f_s}$:} A single-channel DNN is used to learn the source prediction  function $f_s$. For the single-channel DNN architecture, a batch normalization layer used to perform a regularization on the segments in the source dataset $X_s$ to address the issue of the heterogeneity of the source data. Assuming the single-channel DNN contains $H \epsilon N^{*}$ hidden layers, we denote the weight matrix and bias vector of the $j^{th}$ layer $(1 \leq j \leq  H)$ as $W_j$ and $B_j$, respectively. Finally, a softmax layer with $C$ neurons is added, with each neuron of the layer outputting a value that estimates of probability to its corresponding class. This way, the single-channel DNN can classify the segments of source dataset $X_s$ using the source labels $Y_s$.
     \vspace{2.5pt}
     \item \textbf{Initialization of a multi-channel DNN}: A multi-channel DNN is defined to learn the target prediction  function $f_t$. It is trained using target dataset $X_t$ which contains multi-channel segments $X \epsilon \mathbb{R^{L\times S_n}}$, with $S_n$ being the number of channels of the target dataset $X_t$ and target label $Y_t$ which contains associated labels $Y \epsilon \{1, 2,..., C_t\}$ with $C_t$ being the number of classes of the target problem. For the multi-channel DNN architecture, a batch normalization layer is applied to the segments to perform an operation akin to a standard normalization on the input of the network. The $S_n$ sensor channels are then separated. The $k^{th}$ sensor channel $(1 \leq k \leq S_n)$ is processed by an ensemble of hidden layers of the same number and type as the hidden layers of the single DNN. We refer to this ensemble of layers as a branch of the multi-channel DNN. The output of each branch is then concatenated and connected to fully-connected layers. A softmax layer with the target class $C_t$ neurons is added to output class probabilities for the $C_t$ target classes.
     \item \textbf{Transfer of weights from single to multi-channel DNN}: The weights $W_j$ and biases $B_j$ of the $H$ hidden layers of the single DNN learned on $\{D_s, T_s\}$ are transferred to the branches of the multi-channel DNN. In other words, the $j^{th}$ layer of the $k^{th}$ branch (for $1 \leq j \leq H$ and $1 \leq k \leq S_n)$ has its weight and bias matrices $W_k^{(s)}$ and $b_k^{(s)}$ initialized as $W_k$ and $b_k$, respectively.
     \vspace{4pt}
     \item \textbf{Learning of target prediction  function $\mathbf{f_t}$}: The multi-channel DNN is fined-tuned using $(X_t, Y_t)$ to learn $f_t$, which is the predictive function for the target ubiquitous computing problem.
     
\end{itemize}

\subsection{Proposed Architecture}

\begin{figure}[htp]
    \centering
    \includegraphics[width=3.12in]{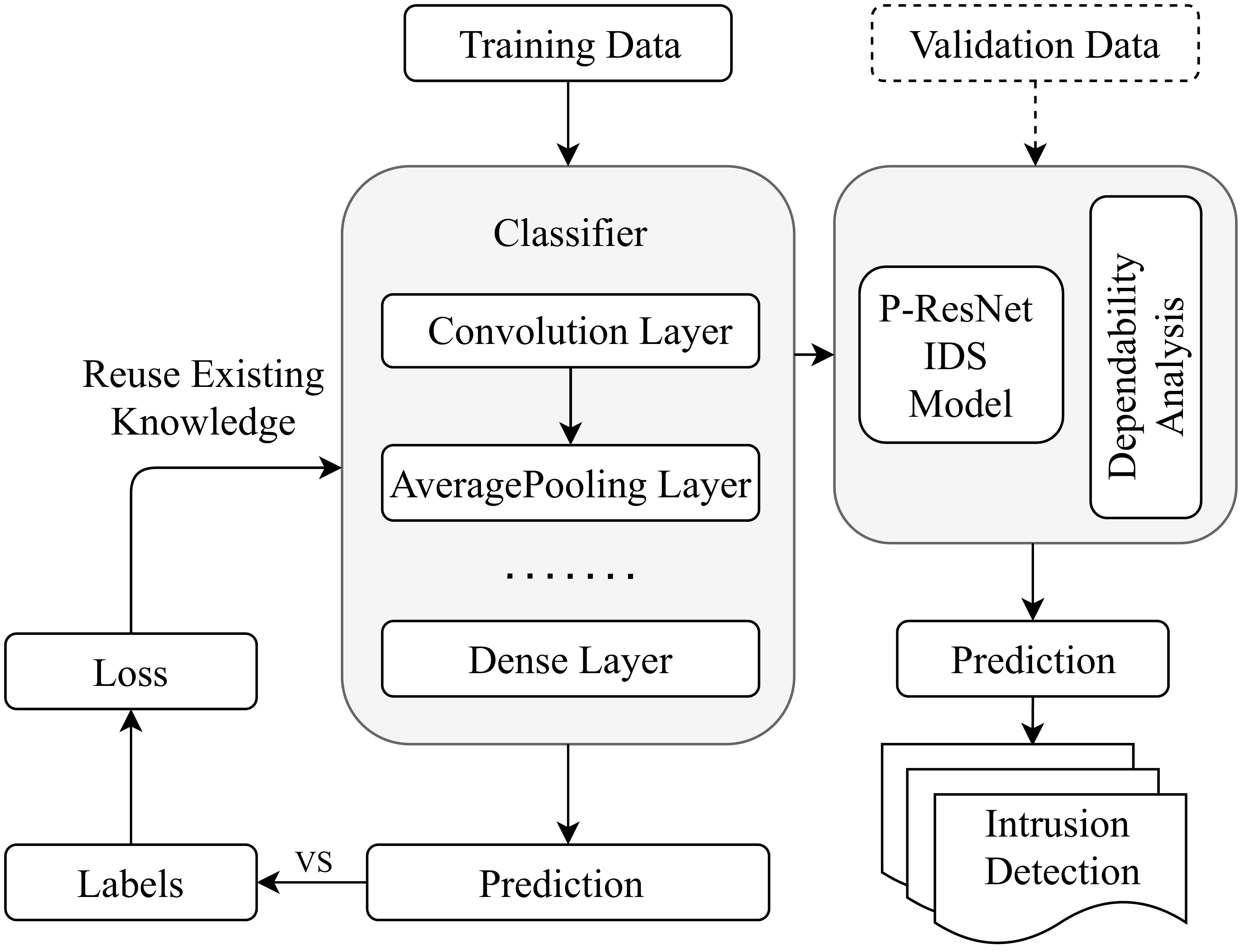}
    \caption{The block diagram of the proposed P-ResNet model}
    \label{The block diagram of the proposed P-ResNet based IDS model}
\end{figure}

Figure \ref{The block diagram of the proposed P-ResNet based IDS model} shows the block diagram of the proposed model for IDS, which predominantly includes two parts: the first is the model training part, and the second is the intrusion detection part. For the training part, after prepossessing the network data, we applied it to our proposed model. 

The most significant parameters of this model have been determined through subsequent empirical experiments. The model with the optimal prediction performance on the training set has been selected as the final intrusion detection model for heterogeneous IoT applications. For the intrusion detection part, we have trained the P-ResNet model with a randomly selected training dataset and validated the model with a validation dataset. The detection performance of the models under the discrete types of parameters is compared. We have also analyzed the dependability performance of our proposed model. Finally, the optimal dependable model has been selected as the final detection model in the field of heterogeneous IoT applications. Table \ref{The hyperparameters of the proposed model} shows the hyper-parameters of the proposed model.

\begin{table}[htp]
\renewcommand{\arraystretch}{1.2}
  \caption{The hyperparameters of the proposed model}
  \label{The hyperparameters of the proposed model}
  \centering
\begin{tabular}{|l|c|}
  \hline
    Hyper-parameters &  Value/Function\\
    \hline
    Number of Hidden Layers  & 4\\
    \hline
    Units in hidden layers & 64, 128, 256, 128\\
    \hline
    Batch size & 64\\
    \hline
    Epochs & 200\\
    \hline
    Hidden layer activation function & ReLu\\
    \hline
    Output layer activation function & Softmax\\
    \hline
    Dropout  & N/A\\
    \hline
    Optimizer & Ada Delta\\
    \hline
    Loss function & Categorical Crossentropy\\
    \hline
\end{tabular}
\end{table}

\begin{figure}[htp]
    \centering
    \includegraphics[width=\columnwidth]{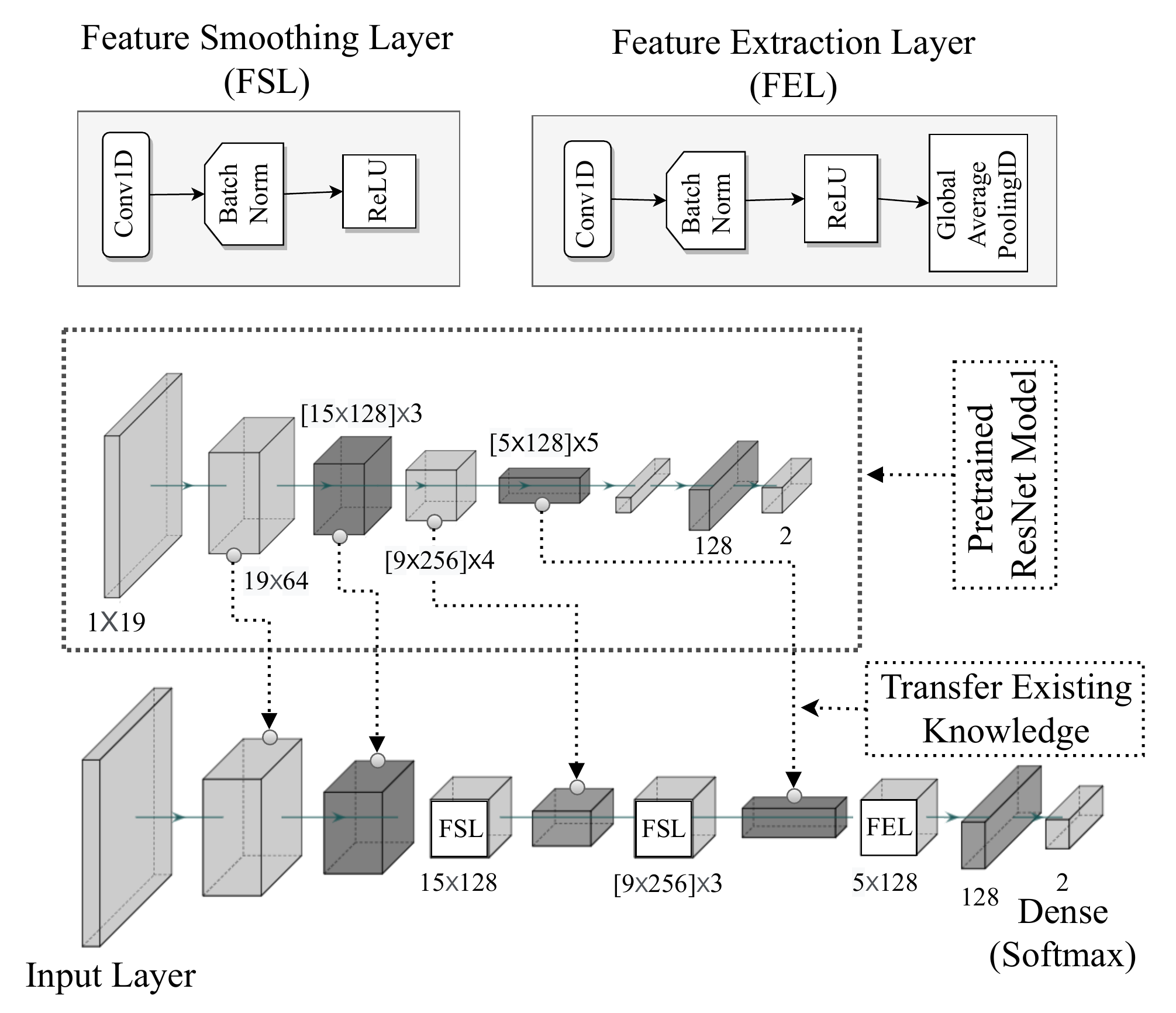}
    \caption{The structure of the proposed P-ResNet model}
    \label{The structure of the proposed P-ResNet model}
\end{figure}

The network architecture of the proposed P-ResNet model is shown in Figure \ref{The structure of the proposed P-ResNet model}. The input of the network is the same shape. There have been four layers in our proposed model without the input layer. The first layer is the combination of 64 filters of kernel length 8, the second layer is the combination of 128 filters of kernel length 8, the third layer is the combination of 256 filters of kernel length 5, and the last layer is the combination of 128 filters of kernel length 3. Each layer is a one-dimensional convolution layer is followed by a \textit{BatchNormalization()} and \textit{Rectified Linear-Unit (ReLU)} activation function, which is called the feature smoothing layer (FSL). FSL has been used to keep the tensors' shape and smooth the pre-trained tensors to learn features from the new training dataset. After the pre-trained residual blocks (transfer pre-trained knowledge), the FSL has been inserted. The feature extraction layer (FEL) used the first three operations to double the number of channels and fatten the feature map into one dimension. The FEL extracted operation-based features and abstracted important information to describe the previous layer, which improves the final fully-connected layer with these abstracted features. 

Each block takes the previous block's outputs as inputs for the current block and performs some non-linearity operations to transform it into a multivariate series whose dimensions are defined by the number of filters in each layer. The fifth layer is the combination of a \textit{GlobalAveragePooling1D()} operation, which takes the input of the previous block and averages each series. This operation drastically reduces the number of parameters in a deep model while enabling the use of a class activation map \cite{Da-9} which allows an interpretation of the learned features. The output of the gap layer is then fed to a \textit{softmax} classification layer whose number of neurons is equal to the number of classes in the dataset. To reduce the overfitting problem of the proposed model, we have considered various strategies, including data augmentation to the training set, adding Gaussian noise during training time, decreasing the network size, and the EarlyStopping method. Skip connections enable feature re-usability and stabilize training and convergence. Finally, we have deployed a skip connections technique of our proposed model by using a vector addition method. Then the gradient would be multiplied by one and its value, which was maintained in the earlier layers. This enabled us to identify the effect of deep transfer learning.

\section{Model Implementation and Evaluation}
\label{sec: model implementation and evaluation}

We have performed all the analyses on a computer with an Intel (R) Core (TM) i7-6800K CPU (3.40GHz) and a GeForce GTX 1080Ti. This section has performed a series of operations on the original combined dataset without changing its properties. Also, we have thoroughly explained the training process and provided an overview of selected models.

\subsection{Dataset Description}

The dataset has been generated from various heterogeneous sources, including seven IoT sensors, (e.g., \textit{fridge\_sensor}, \textit{GPS\_tracker\_sensor}, \textit{motion\_light\_sensor}, \textit{garage\_door\_sensor}, \textit{modbus\_sensor}, \textit{thermostat\_sensor}, and \textit{weather\_sensors}). The \textit{fridge\_sensor} measures the fridge\_temperature and temperature\_condition, whether it is high or low, based on a preset threshold value. The \textit{GPS\_tracker\_sensor} determines the geographic position, such as latitude and longitude of an object. The \textit{motion\_light\_sensor} includes two characteristics, i. e., motion\_status and light\_status. The light\_status is changed based on the pseudo-random signal generated by the motion\_status. The \textit{garage\_door\_sensor} observed two-states, door\_state and sphone\_signal. The first is observed when the door is open or closed and the second is where the signal is true or false when receiving the door signal on a phone. The \textit{modbus\_sensor} simulates the functionality of the Modbus devices for communication with each other with the help of some register (e.g., input\_register, discrete\_register, holding\_register, and a coil\_over\_register). The \textit{thermostat\_sensor} governs the temperature of a physical system by regulating a heating or cooling system, where current\_temperature and thermostat\_status are the fundamental features. The \textit{weather\_sensors} produce  air\_temperature, air\_pressure, and air\_humidity data. 

\begin{figure}[htp]
    \centering
    \includegraphics[width=3.1in]{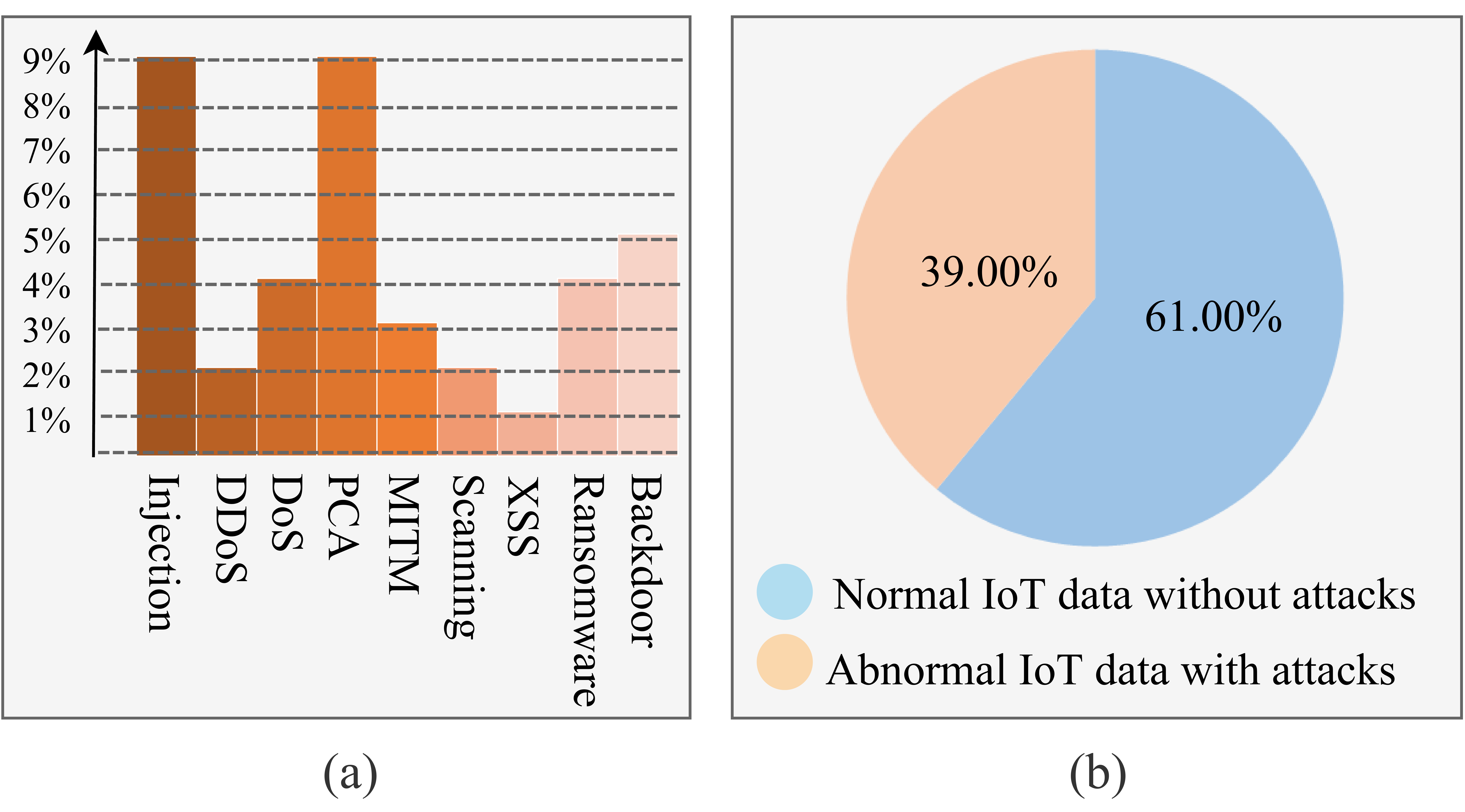}
    \caption{ (a) Statistics of the abnormal dataset with attacks; (b) Class distribution of the combined dataset}
    \label{Statistics of the combined dataset}
\end{figure}

Each IoT dataset has been combined into one CSV file by a Python script. The combined dataset has 22 features, including the label feature. The label feature contains two quantitative values, i.e., 0 and 1, which indicate normal and attack scenarios, respectively. Nine types of cyber-attacks (\textit{denial of service (DoS)}, \textit{distributed denial of service (DDoS)}, \textit{data injection}, \textit{man-in-the-middle (MITM)}, \textit{backdoor}, \textit{password cracking attack (PCA)}, \textit{scanning}, \textit{cross-site scripting (XSS)}, and \textit{ransomware}) have been considered against the IoT sensors mentioned above. Figure \ref{Statistics of the combined dataset}(a) shows the statistics of the abnormal dataset with attacks and figure \ref{Statistics of the combined dataset}(b) shows the class distribution of the combined dataset.

\subsection{Data Preparation}

To achieve optimal performance and improve the learning process, it is necessary to clean and prepare the dataset before applying DTL methods. Data preparation is generally done by removing unnecessary features, checking the variation of independent features, converting non-numerical features, removing outliers, and replacing missing values if they exist. The two fundamental steps apply during the data preparation process. The first is data pre-processing, and the second is data transformation step.

\subsubsection{Data Pre-processing}

This dataset originates from multiple heterogeneous sources. Due to its vast size, this dataset is highly susceptible to missing and noisy data. This section discusses the essential steps in data pre-processing: data-cleaning and data-integration. 

\begin{itemize}
  \item \textbf{Data Cleaning:} First, we have applied various techniques to remove noise and clean inconsistencies in the dataset. For example, \textit{Rosner's Test} for outliers checking, and the \textit{predictive mean matching} method for imputing missing values. Then, to apply DTL models, we have converted the qualitative values into quantitative values. To cite an example, the \textit{door\_state} feature in the dataset, which has qualitative values of ‘open’ and ‘closed’, has been converted into ‘1’ and ‘0’. The conversion to quantitative values was performed using a numerical convolution label-encoding library \textit{numconv} \cite{Geron-56}. Some features, such as - date, time, and timestamp, have been omitted from feature vectors as they may cause them to overfit the training data. Furthermore, the input data shape has been reshaped into three dimensions to feed the models by applying \textit{numpy.reshape} mechanism with \textit{swapaxes} and \textit{concatenate} methods.
  
  \item \textbf{Data Integration:} To improve the accuracy and speed of the training and validation processes, the data integration technique helped us  reduce and avoid redundancies in the resulting dataset. This dataset originates from multiple heterogeneous sources. So, it is essential to combine all the IoT sensors' data by \textit{redundancy} and \textit{correlation} analysis. This analysis has measured how strongly one feature, i.e., \textit{door\_state} implies the other, i.e., \textit{light\_status}. Figure \ref{Correlation visualization} shows the correlation between different features. For our analysis, we have evaluated the correlation between all the features using the following \textit{Pearson’s product-moment coefficient} equation.
  
\begin{align}
  C_{orrelation} =
  \frac{ \sum_{i=1}^{n}(x_i-\bar{x})(y_i-\bar{y})}{
        \sqrt{\sum_{i=1}^{n}(x_i-\bar{x})^2}\sqrt{\sum_{i=1}^{n}(y_i-\bar{y})^2}}
\end{align}

Where \emph{n} is the number of tuples, \emph{x\textsubscript{i}} and \emph{y\textsubscript{i}} are the respective values in tuple \emph{i}, and $\bar{x}$ and $\bar{y}$ are the respective mean values of \emph{x} and \emph{y}.

\begin{figure}[htp]
    \centering
    \includegraphics[width=\columnwidth]{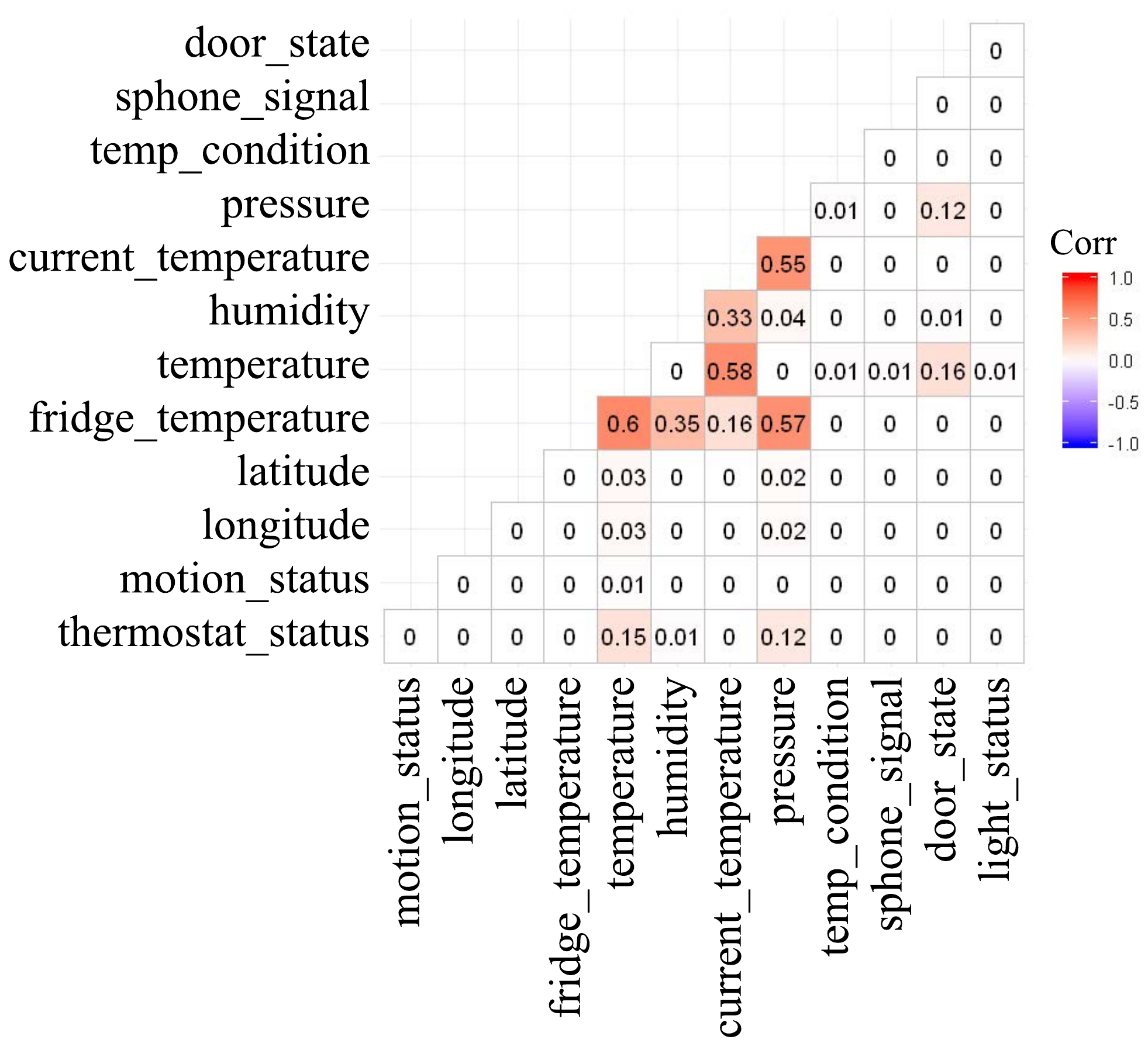}
    \caption{Correlation visualization}
    \label{Correlation visualization}
\end{figure}
  
\end{itemize}

\subsubsection{Data Transformation}

We applied this processing step to achieve more efficient resulting processes and easily understand the patterns. Some selected features have larger values than others, which leads to incorrect performance, though some models might prefer the large feature values. We have performed these strategies to scale the selected feature values within a range between [0.0, 1.0] without changing the characteristics of the data \cite{Aminanto-58}. As shown in the following equation~\eqref{eq4}, a  technique called \textit{minimum-maximum normalization} has been used to scale the selected feature values within the range.

\begin{align}
\label{eq4}
      N_{ormalized}V_{alue} = \frac{(X-X_{min})} {(X_{max}- X_{min})}  
\end{align}

Where, X is an original value and X\textsubscript{max} and X\textsubscript{min} are the maximum and minimum values of the feature, respectively.

\subsection{Training Process}

As discussed in the previous section, all the datasets have been converted into a combined dataset by a Python script containing the training and validation data. First of all, we have divided the combined dataset into the training data set (80\%) and the test data set (20\%) by using the \textit{train\_test\_split} method of the \textit{scikit\_learn} library. The training data set has been used to train the selected models. On the other hand, the test data set has been used to further assess the trained classifier of those models. 

\begin{figure}[htp]
    \centering
    \includegraphics[width=\columnwidth]{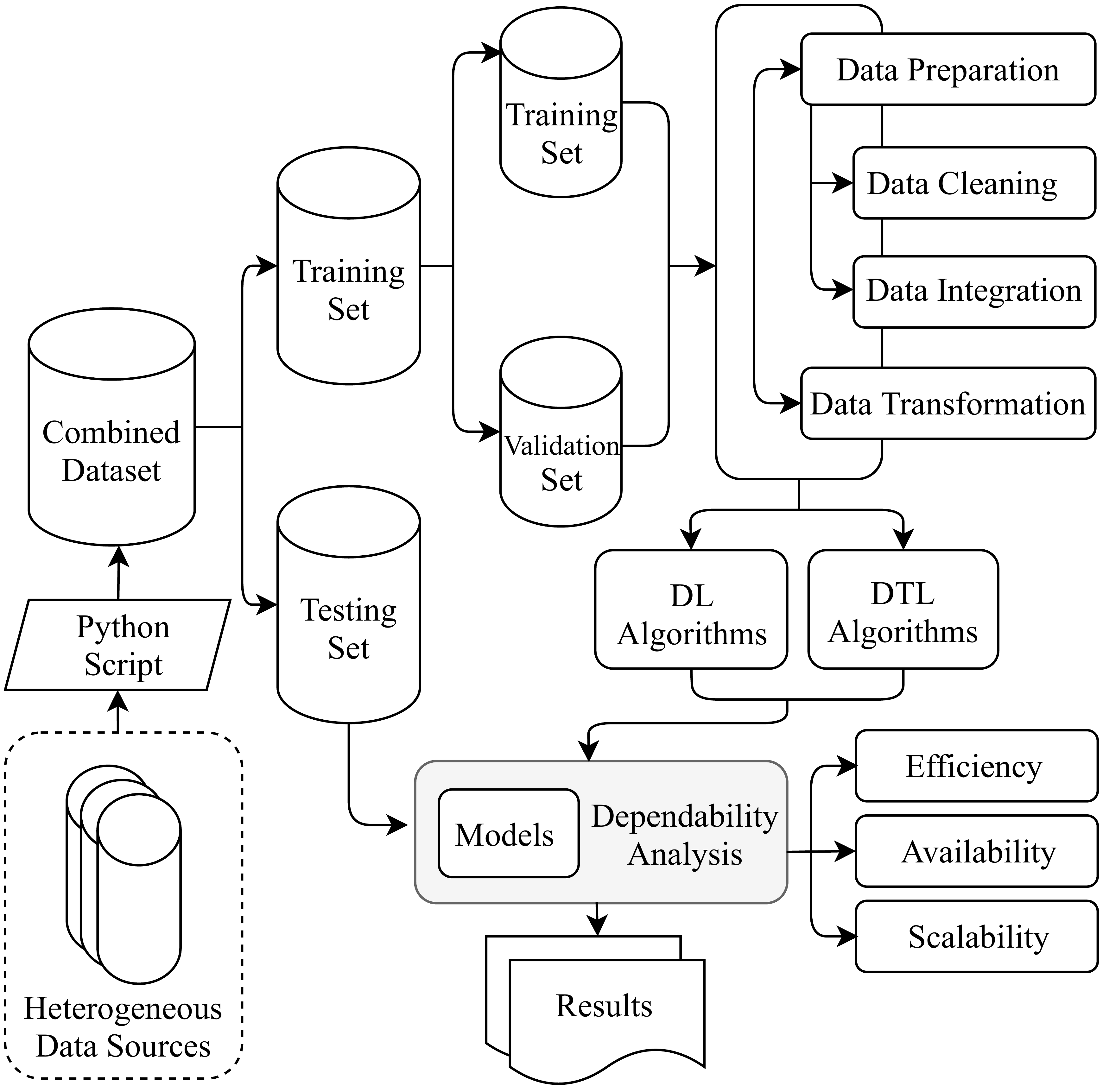}
    \caption{Overall evaluation process of the selected models}
    \label{Figure: Evaluation process of our idea}
\end{figure}

Furthermore, we again split the training data (80\% of the total data) into the new training data (80\%) for training the selected model and validation data (20\%) for hyperparameters’ optimization. Proportions of 80\% for the training dataset and 20\% for the validation dataset have been chosen as suggested in \cite{Geron-56}. To avoid the overfitting problem, this splitting ratio has been considered the best ratio between the training and the validation dataset \cite{Guyon-60}. We want to highlight some DTL models that give better performance than others \cite{Da-9}. We have selected Fully Convolutional Networks (FCN), LeCun Network (LeNet), Inception Network (IncepNet), Multi Channel Deep Convolutional Neural Network (MCDCNN), Convolutional Neural Network (CNN), Long Short-Term Memory (LSTM), Multilayer Perceptron (MLP), and our proposed Residual Neural Network (P-ResNet). We have also considered DL algorithms because some of their variants have been successfully applied to solve classification tasks related to intrusion detection \cite{Nisioti-67, Omar-72}. Therefore, we have considered Long Short-Term Memory (LSTM), Neural Network (NN), Convolutional Neural Network (CNN), and Recurrent Neural Network (RNN) algorithms because of their optimal performance. The overall performance of the selected models has been evaluated with a wide range of tested hyperparameters. Figure \ref{Figure: Evaluation process of our idea} summarizes the steps involved in evaluating the performances of the selected models. 

\section{Result Analysis}
\label{sec: result analysis}

This section discusses the overall performance of the selected models. We incorporated a wide range of analysis scenarios with varying measurement indicators, including accuracy, precision, recall, ROC AUC, and f1-score. We also concentrated on the optimization of the hyper-parameters for improved performance.

\subsection{DL Metrics Analysis}

\begin{figure}[htp]
    \centering
    \includegraphics[width=\columnwidth]{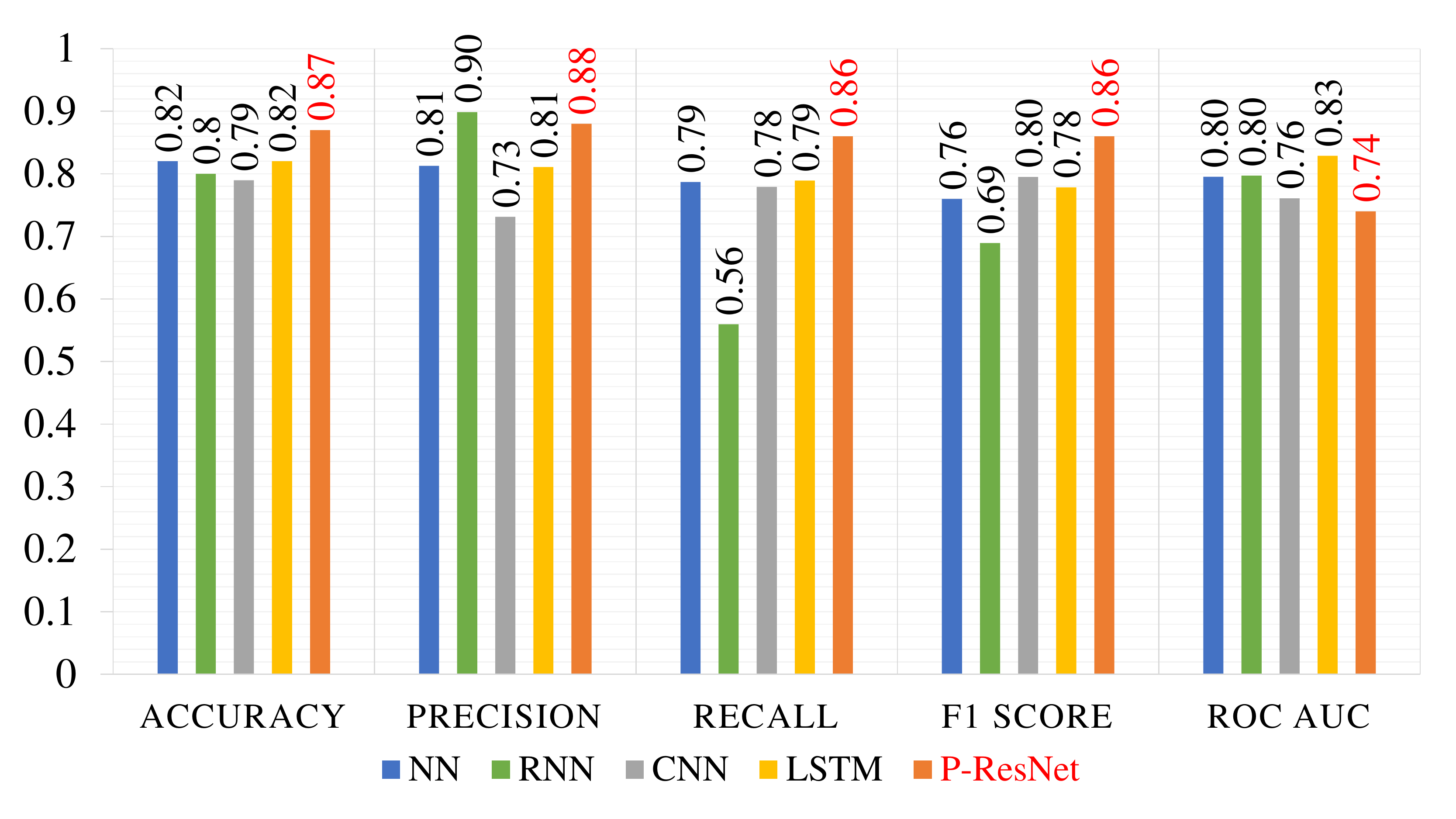}
    \caption{Performance comparison of DL algorithms' with the proposed DTL-based P-ResNet model}
    \label{DL algorithms' performance visualization}
\end{figure}

In recent years, DL algorithms have advanced significantly, and some of the variants of DL algorithms have been successfully applied to solve classification tasks related to intrusion detection \cite{Nisioti-67}. Therefore, we considered DL algorithms in this section because of their optimal performance. In this experiment, we considered LSTM, NN, CNN, and RNN. The LSTM and NN models show the highest accuracy score of 0.82. Dramatically, the other performance metrics of these two models are almost similar. Figure \ref{DL algorithms' performance visualization} shows the performance comparison between selected DL models and the proposed DTL-based P-ResNet model. For the LSTM model, we used three hidden-layers, where the units of hidden layers were 128, 100, and 64, respectively. The \textit{tanh} is the hidden layer activation function, and \textit{Adam} is used as an optimizer. Also, \textit{sigmoid} is used as a network output activation function, and ``categorical\texttt{\_}crossentropy" is used as a loss function. The RNN and CNN models outperform most TML algorithms, with an accuracy score of 0.80 and 0.79, respectively. Furthermore, both the NN and CNN models used the same optimizer and activation function but different types of the hidden layer activation function. Particularly for the RNN model, the \textit{softmax} is used as the network output activation function, and ``categorical\texttt{\_}crossentropy" is used as the loss function. On the other hand, for the CNN model, the number of hidden layers is four and ``binary\texttt{\_}crossentropy" is used as the loss function. Finally, considering all the significant parameters of the DL models, the NN and LSTM models have shown the optimal performance, which indicates that most of the predicted labels are correctly classified. However, when we applied our proposed P-ResNet model to the same dataset, the model performed significantly better in most cases. In such cases, accuracy, precision, f1-score, and ROC AUC are better than in other experimental scenarios.

\subsection{DTL Metrics Analysis}

In recent years, the progression of deep transfer learning and its great success in different fields mean it has emerged as a potential solution for dependable network intrusion detection \cite{Nisioti-67}. Besides considering the fundamental evaluation criteria, we also incorporated other deep learning evaluation criteria, for example, the numbers of hidden layers, units in the hidden layers, output layer activation functions, loss functions, etc. These criteria helped to calculate and report the classification results effectively. First, we consider the quantitative performance of DTL models. Table \ref{Table: DTL algorithms' performance comparison metrics} shows the quantitative performance summary of the DTL models.

\begin{table}[htp]
\renewcommand{\arraystretch}{1.3}
\caption{DTL algorithms' performance comparison metrics}
\label{Table: DTL algorithms' performance comparison metrics}
\centering
\begin{tabular}{cccccc}
\hline
Algorithm & Accuracy & Precision & Recall & F1Score & ROC AUC\\
\hline
FCN & 0.84 & 0.85 & 0.84 & 0.83 & 0.81\\
\hline
LeNet & 0.80 & 0.82 & 0.80 & 0.79 &0.76\\
\hline
IncepNet & 0.80 & 0.86 & 0.80 & 0.81 & 0.73\\
\hline
MCDCNN & 0.80 & 0.83 & 0.80 & 0.79 & 0.76\\
\hline
CNN & 0.81 & 0.83 & 0.81 & 0.80 & 0.76\\
\hline
LSTM & 0.85 & 0.84 & 0.77 & 0.77 & 0.83\\
\hline
MLP & 0.73 & 0.74 & 0.78 & 0.77 & 0.74\\
\hline
P-ResNet & 0.87 & 0.88 & 0.86 & 0.86 & 0.83\\
\hline
\end{tabular}
\end{table}

\begin{figure}[htp]
    \centering
    \includegraphics[width=\columnwidth]{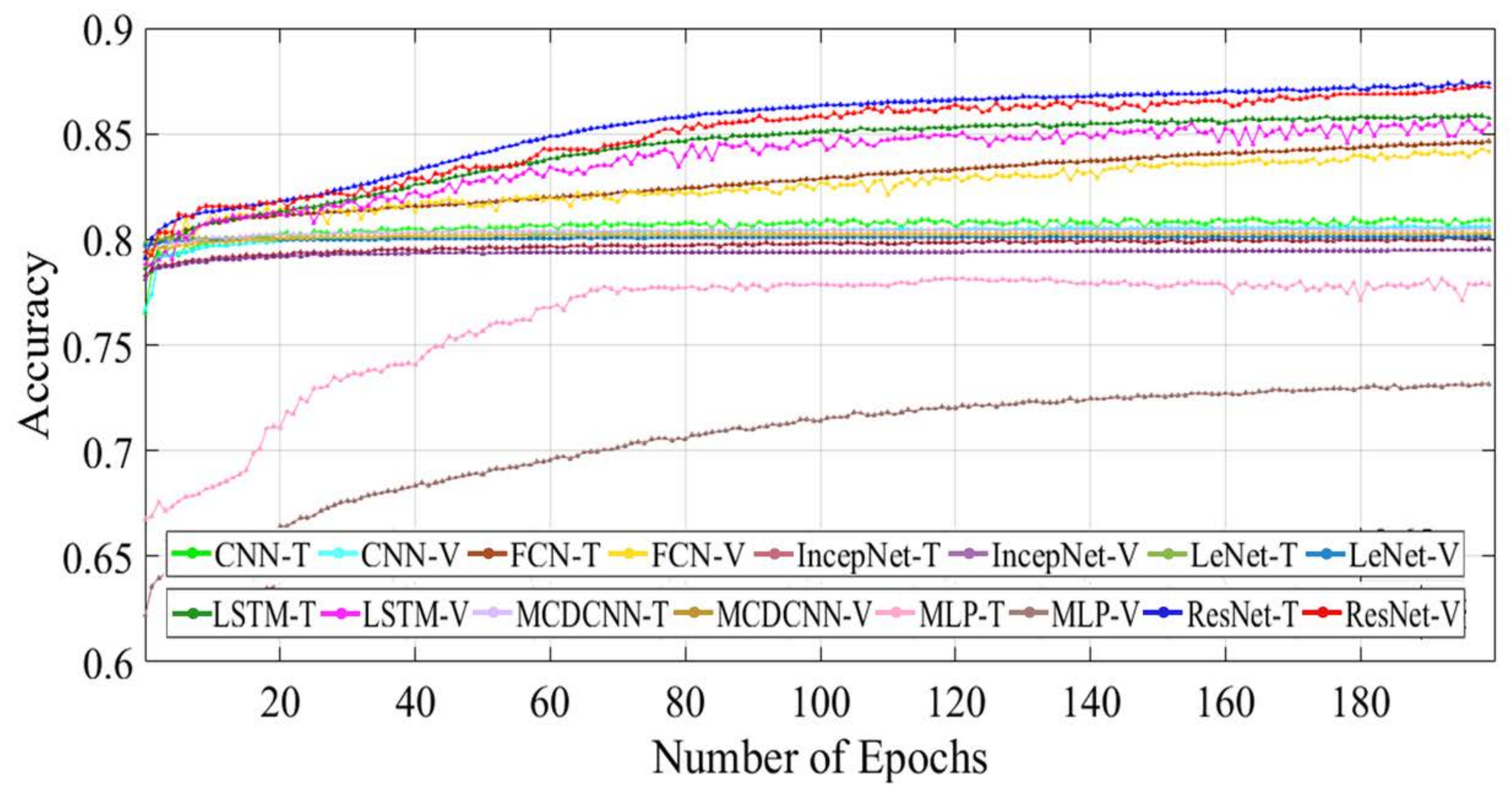}
    \caption{DTL algorithms' training and validation accuracy}
    \label{DTL algorithms' training and validation accuracy}
\end{figure}

The proposed P-ResNet model shows optimal performance compared to the others, with an accuracy score of 0.87, precision score of 0.88, recall score of 0.86, f1-score of 0.86, and ROC AUC score of 0.83. We used four hidden-layers in this model, where \textit{ReLu} is the hidden layer activation function. Also, \textit{softmax} is used as a network output activation function, and ``categorical\texttt{\_}crossentropy" is used as a loss function along with the \textit{Adam} optimizer. Figure \ref{DTL algorithms' training and validation accuracy} shows the accuracy score of every single epoch for both the training and validation phases of the selected models. Considering epoch numbers 170 to 200 for both phases, the flattening characteristics of the curve and accuracy are not increasing literally. Therefore, we have considered 200 epochs for our analysis. The proposed model shows the highest accuracy score in both the training and validation phases, whereas, the MLP model shows the lowest accuracy score during any setting of the epoch number within 1 to 200 in both the training and validation phases. 

In detail, according to Figure \ref{DTL algorithms' training and validation accuracy}, the accuracy of the MLP model starts around 0.67 for the training phase and 0.63 for the validation phase in epoch number 10. However, it increased dramatically to around 0.78 in epoch number 65 and 0.74 in epoch number 140 for the training and validation phases, respectively. However, for the training phase, the accuracy score (0.78) remains stable in epoch number 66 to 200. 

\begin{figure}[htp]
    \centering
    \includegraphics[width=\columnwidth]{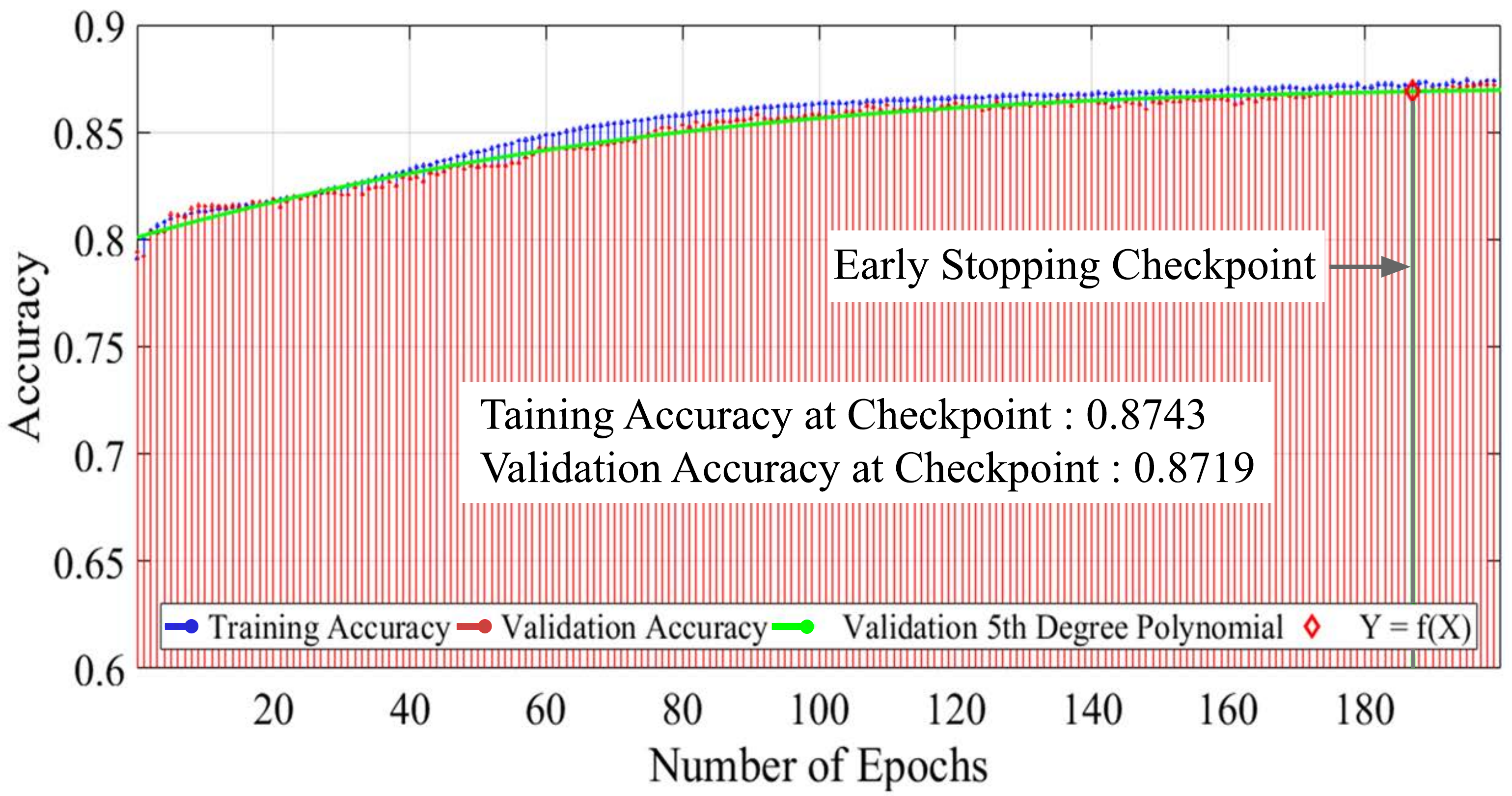}
    \caption{Training and validation accuracy of proposed model}
    \label{Training and validation accuracy of proposed P-ResNet model}
\end{figure}

On the other hand, for the validation phase, the accuracy score (0.74) remains stable between epoch numbers 141 and 200. As shown in Figure \ref{DTL algorithms' training and validation accuracy}, the training and validation accuracy of CNN, IncepNet, LeNet, and MCDCNN models remains steady between epochs 1 and 200. The accuracy of LSTM and FCN models starts with a score of 0.82. However, this score rises gradually with the increase of the epoch number and reaches approximately 0.86 when the epoch number is 160 and then remains stable between epoch numbers 161 and 200 for both phases. The remarkable point is that the behavior of the training phase is almost identical to the validation phase. To better understand our proposed model Figure \ref{Training and validation accuracy of proposed P-ResNet model} shows the trend of the accuracy score of both phases. The proposed model's accuracy increases rapidly at epoch number 60, reaching a peak of close to 0.87 at epoch number 169. However, as shown in Figure \ref{Training and validation accuracy of proposed P-ResNet model}, it remains almost stable up to the early stopping checkpoint.

\begin{figure}[htp]
    \centering
    \includegraphics[width=\columnwidth]{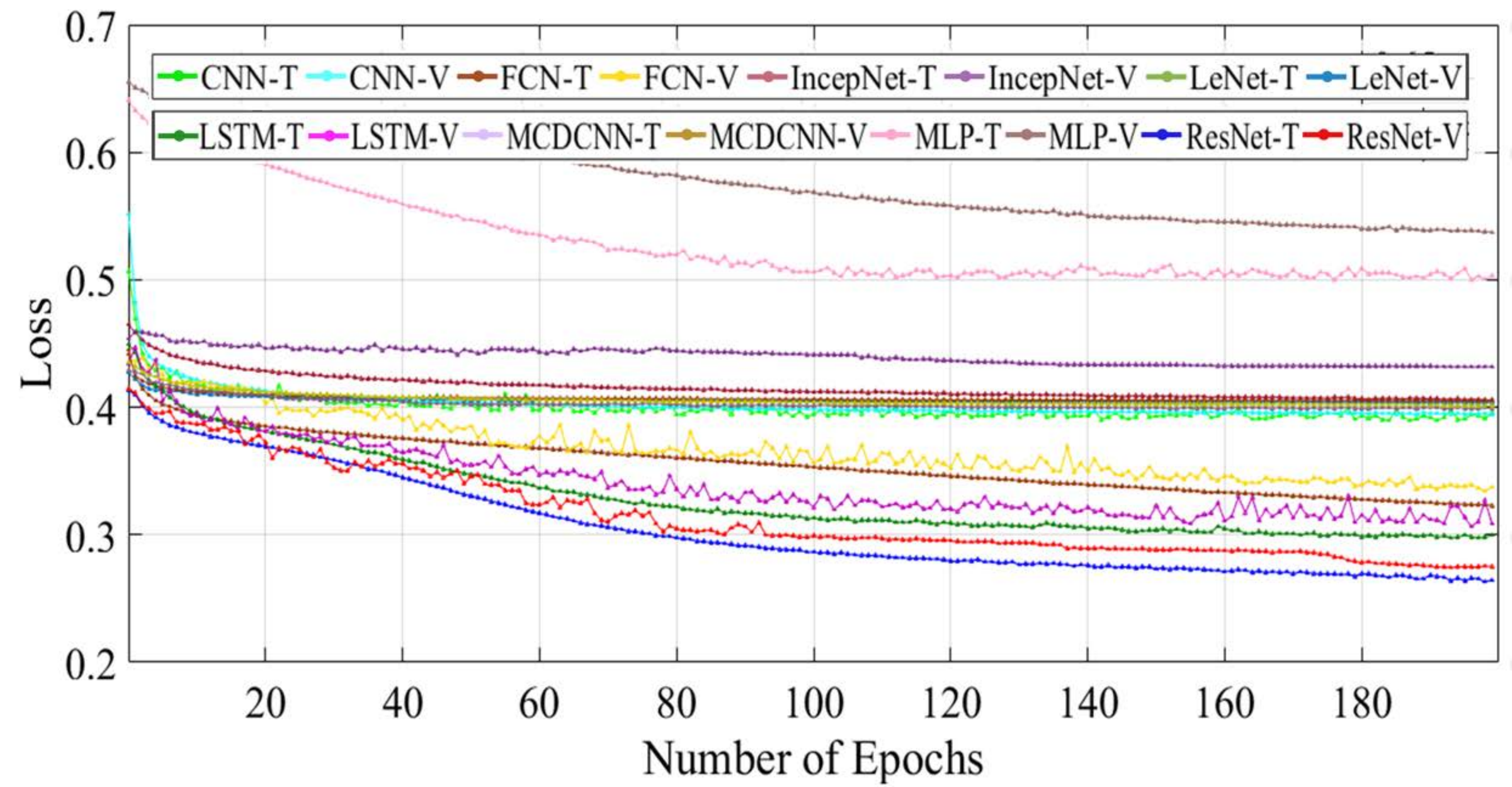}
    \caption{DTL algorithms' training and validation loss}
    \label{DTL algorithms' training and validation loss}
\end{figure}

\begin{figure}[htp]
    \centering
    \includegraphics[width=\columnwidth]{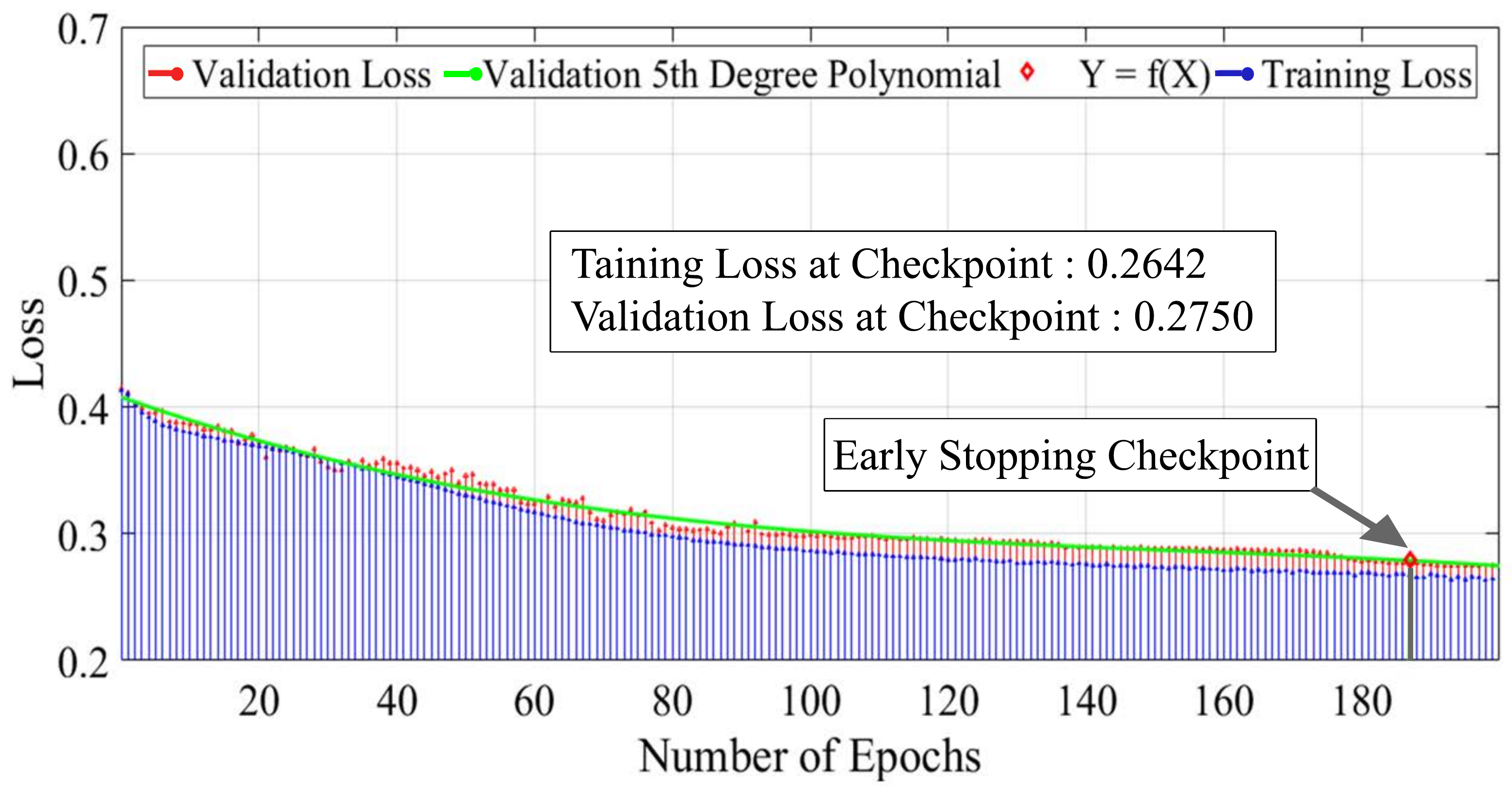}
    \caption{Training and validation loss of proposed model}
    \label{Training and validation loss of proposed P-ResNet model}
\end{figure}

Next, we analyzed the losses of each model. The training and validation losses of every epoch are shown in Figure \ref{DTL algorithms' training and validation loss}. The MLP model shows the highest losses in the training and validation phases. The highest loss of this model indicates that the model cannot provide a reliable classification between normal and attack scenarios. On the other hand, our proposed model shows the lowest loss for any epoch setting within 1 to 200. In detail, the loss of the proposed model starts around 0.41 for the training phase and 0.42 for the validation phase, as shown in Figure \ref{Training and validation loss of proposed P-ResNet model}. However, it decreases to 0.27 for the epoch number 140 and 0.28 when the number is 145 for the training and validation phases, respectively. The loss score (0.27) remains stable between epoch number 141 and the early stopping checkpoint, which has been indicated by the \textit{red circle} in Figure \ref{Training and validation loss of proposed P-ResNet model}. The losses of the CNN, IncepNet, LeNet, and MCDCNN models remain almost steady during both the training and validation phases, between epochs 1 and 200, which is shown in Figure \ref{DTL algorithms' training and validation loss}. The loss of the LSTM and FCN models is almost 0.42 at the beginning, which declines gradually to approximately 0.31 at epoch number 140 and remains stable for both phases.

During a concept drift scenario, the learning model drifts, which can be understood as changes in the relationship between the input and target output. In an IDS setup, their are significant challenges arising from the `concept drift', especially in the context of streaming IoT data, where the dataset may not always be static but rather an evolving and moving target for detection. Authors in~\cite{Andresini-68} highlight a set of open challenges faced by the modern ML-based IDS technologies and emphasize adopting concept drift within the solution framework. Gama et al.~\cite{Gama-70} highlight that transfer learning between the different contexts can be a promising solution to deal with the `concept drift' challenges. McKay et al.~\cite{McKay-71} recently implemented an online transfer learning-based solution for concept drifting data streams. Hence, with confidence, we can argue that the proposed transfer learning-based solution has the potential to deal with concept drifting contexts, which we have scoped for future work.

\subsection{Performance Analysis on Training and Testing Time}

The training time, testing time, and number of parameters were also calculated for each DL and DTL model. Table~\ref{Table: DTL algorithms' testing and training time} compares the total parameters, training time, and testing time for each model. In terms of the DL models, NN requires 2.660 seconds for a total of 50,985 trainable parameters, which is less testing time and a lower number of parameters than other DL models. Moreover, the NN model can be explained as other DL models were lazy learners who used the training phase to store the data and then used the data during the test phase to make a prediction, which makes the testing phase slower, though the number of parameters (trainable) was slightly increasing.

\begin{table}[htp]
\renewcommand{\arraystretch}{1.3}
\caption{Comparison of total parameters, training time, testing time of DL and DTL algorithms}
\label{Table: DTL algorithms' testing and training time}
\centering
\begin{tabular}{ccccc}
\hline
& Algorithm & Params & Training Time (s) & Testing Time (s)\\
\hline
{\multirow{4}{*}{\rotatebox{90}{\textbf{Deep Learn.}}}} 
& NN & 50,985 & 3242.180 & 2.660 \\
\cline{2-5}
& RNN & 59,205 & 3258.245 & 3.105 \\
\cline{2-5}
& CNN & 56,086 & 2763.788 & 2.667\\
\cline{2-5}
& LSTM & 66,689 & 25921.078 & 5.940\\
\hline
{\multirow{8}{*}{\rotatebox{90}{\textbf{Deep Transfer Learning}}}} 
& FCN & 265,986 & 46756.338 & 7.722 \\
\cline{2-5}
& LeNet & 260,052 & 38936.911 & 5.690 \\
\cline{2-5}
& IncepNet & 233,074 & 36657.213 & 4.229\\
\cline{2-5}
& MCDCNN & 522,998 & 34198.530 & 3.075\\
\cline{2-5}
& CNN & 202,686 & 33869.838 & 2.718\\
\cline{2-5}
& LSTM & 385,538 & 44859.869 & 5.712\\
\cline{2-5}
& MLP & 513,002 & 27065.069 & 3.192 \\
\cline{2-5}
& P-ResNet & 506,818 & 24401.586 & 3.014 \\
\hline
\end{tabular}
\end{table}

On the other hand, in terms of the training and testing time, the FCN model requires the longest training and testing time of 46756.338 seconds and 7.722 seconds, respectively, for a total of 265,986 trainable parameters. Surprisingly, the CNN model has the lowest testing time of 2.718 seconds and training time of 33869.838 seconds, the third-lowest among all the DTL models. Furthermore, the LSTM model has a testing time of 5.721 seconds and training time of 44859.869 seconds, the third-longest testing time after the LeNet model, which has a training time of 38936.911 seconds and testing time of 5.690 seconds. One would be concerned that the LSTM and LeNet models have more parameters than the CNN. Both the MLP and MCDCNN models have almost the same testing time and number of parameters. Our proposed P-ResNet model requires a testing time of 3.014 seconds and training time of 24401.586 seconds, for a total of 511,002 trainable parameters, which is less than most of the selected DTL models. This can be explained as our proposed model is a fast learner that uses the training phase to train efficiently, and then uses the data during the test phase to make a prediction, which makes the testing phase faster. Considering all the performance indicators, we think our proposed model can be used for practical applications to protect Industry 4.0 manufacturing and other critical heterogeneous IoT-based infrastructures where dependable IDS system and secure data processing are the main challenges. 

\subsection{Dependability Performance Analysis}

\begin{figure}[htp]
    \centering
    \includegraphics[width=\columnwidth]{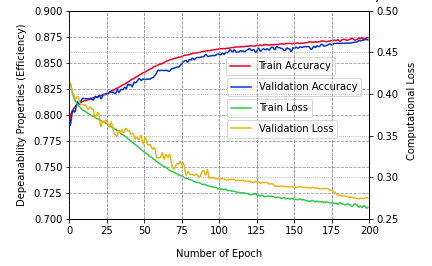}
    \caption{Dependability performance analysis (efficiency) of our proposed P-ResNet model}
    \label{Dependability performance analysis (efficiency) of our proposed P-ResNet model}
\end{figure}

In this section, we explored the dependability performance analysis of our proposed model. Dependability performance analysis includes availability, efficiency, and scalability features~\cite{Bhuiyan-4}. We consider various strategies to select features and then perform our proposed model to accurately classify normal and attack scenarios without any type of failure or undergoing a repair action, which maintains the availability of our proposed model. Moreover, extensive analysis and performance evaluation (e.g., accuracy, precision, ROC AUC, etc.) show that the proposed model is more efficient and demonstrates better performance than several other existing approaches with low computational loss. Figure~\ref{Dependability performance analysis (efficiency) of our proposed P-ResNet model} shows the efficiency and computational loss of the proposed model. 

\begin{figure}[htp]
    \centering
    \includegraphics[width=\columnwidth]{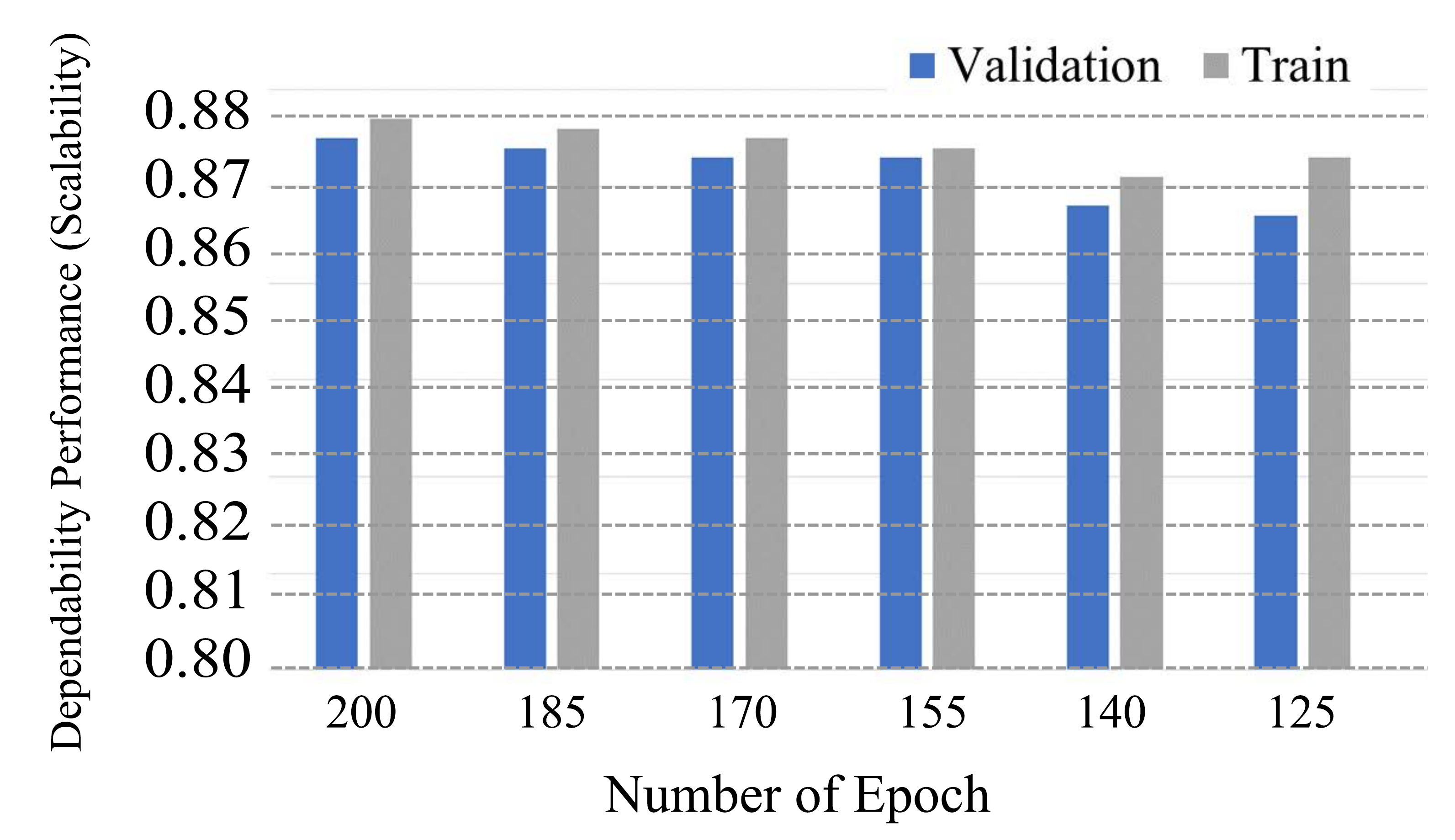}
    \caption{Dependability performance analysis (scalability) of our proposed P-ResNet model}
    \label{Dependability performance analysis (scalability) of our proposed P-ResNet model}
\end{figure}

Finally, we observed an increase in the scalability properties of our proposed model by including various heterogeneous trusted data sources in the training dataset with maximum consistency that were acquired from a wider range of IoT sensors. As a result, our proposed model's accuracy remained almost the same when we increased the epoch number from 125 to 200, indicating its scalability. Figure~\ref{Dependability performance analysis (scalability) of our proposed P-ResNet model} shows the scalability performance of the proposed model.

\section{Conclusion and Future Work}
\label{sec: conclusion and future work}

In this paper, we proposed a deep transfer learning-based dependable intrusion detection model along with improved performance in comparison to several other existing approaches. The overall accuracy of the proposed detection model is 87\%, ensuring dependability and low time complexity. Moreover, the model also greatly improves the precision score of 88\%, the recall score of 86\%, and the f1-score of 86\%, which are higher than the benchmark models. This demonstrates that the proposed P-ResNet model can effectively classify attack instances within various heterogeneous IoT networks, ensuring its dependability. Moreover, the proposed model has proven its potential to efficiently exhibit anomalous data identification to protect Industry 4.0 manufacturing and other critical infrastructures where dependable IoT-enabled automation and secure data processing are the main challenges. More precisely, our proposed mechanism provides a guideline for future research of dependable intrusion detection for IoT networks. In the future, we will concentrate on many IoT sensors and optimize the hyper-parameters of the proposed deep transfer learning for improved performance. We will also try to implement this proposed deep learning model based on privacy-preserving decentralized devices or servers.

\ifCLASSOPTIONcaptionsoff
  \newpage
\fi

\bibliography{sample}
\bibliographystyle{ieeetr}

\begin{IEEEbiography}[{\includegraphics[width=1in,height=1.25in,clip,keepaspectratio]{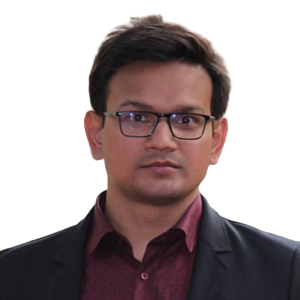}}]{Sk. Tanzir Mehedi} is studying MSc. Engineering degree in Information and Communication Technology at Mawlana Bhashani Science and Technology University, Bangladesh. He worked as a Data Analyst Engineer (Trainee) at Fujitsu Research Institute, Tokyo, Japan, majoring in R and Python programming. He is also a Java and PHP developer and an open web contributor. Currently, he is serving as a Lecturer at the Department of Information Technology (IT), University of Information Technology and Sciences (UITS), Baridhara, Dhaka-1212, Bangladesh. His research interests include data science, blockchain technology, machine learning, deep federated learning, intrusion detection, and data privacy protection.
\end{IEEEbiography}

\begin{IEEEbiography}[{\includegraphics[width=1in,height=1.25in,clip,keepaspectratio]{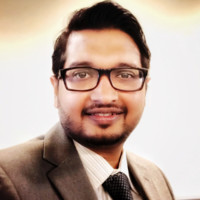}}]{Adnan Anwar} is a Cyber Security academic at Deakin University, and a member of the Centre for Cyber Security Research and Innovation (CSRI). Previously he has worked as a Data Scientist and analytics team leader at Flow Power. He has over 10 years of industrial, research, and teaching experience in universities and research laboratories including NICTA (now, Data61 of CSIRO), University of New South Wales (UNSW), La Trobe University, and Deakin University. He received his PhD and Master by Research degree from UNSW at the Australian Defence Force Academy (ADFA). He has authored over 60+ articles including journals, conference articles and book chapters in prestigious venues (H-index is 17). He has attracted research income from Government, Defence, Industries and received numerous awards at Deakin for excellence in research and teaching. Dr. Anwar’s research has greatly improved the state of the art in artificial intelligence and data-driven cybersecurity research for critical infrastructure in Australia, while his teaching is helping to develop the next generation of Australian experts (over 1200 graduates) in the area of data analytics for security and privacy.
\end{IEEEbiography}

\begin{IEEEbiography}[{\includegraphics[width=1in,height=1.25in,clip,keepaspectratio]{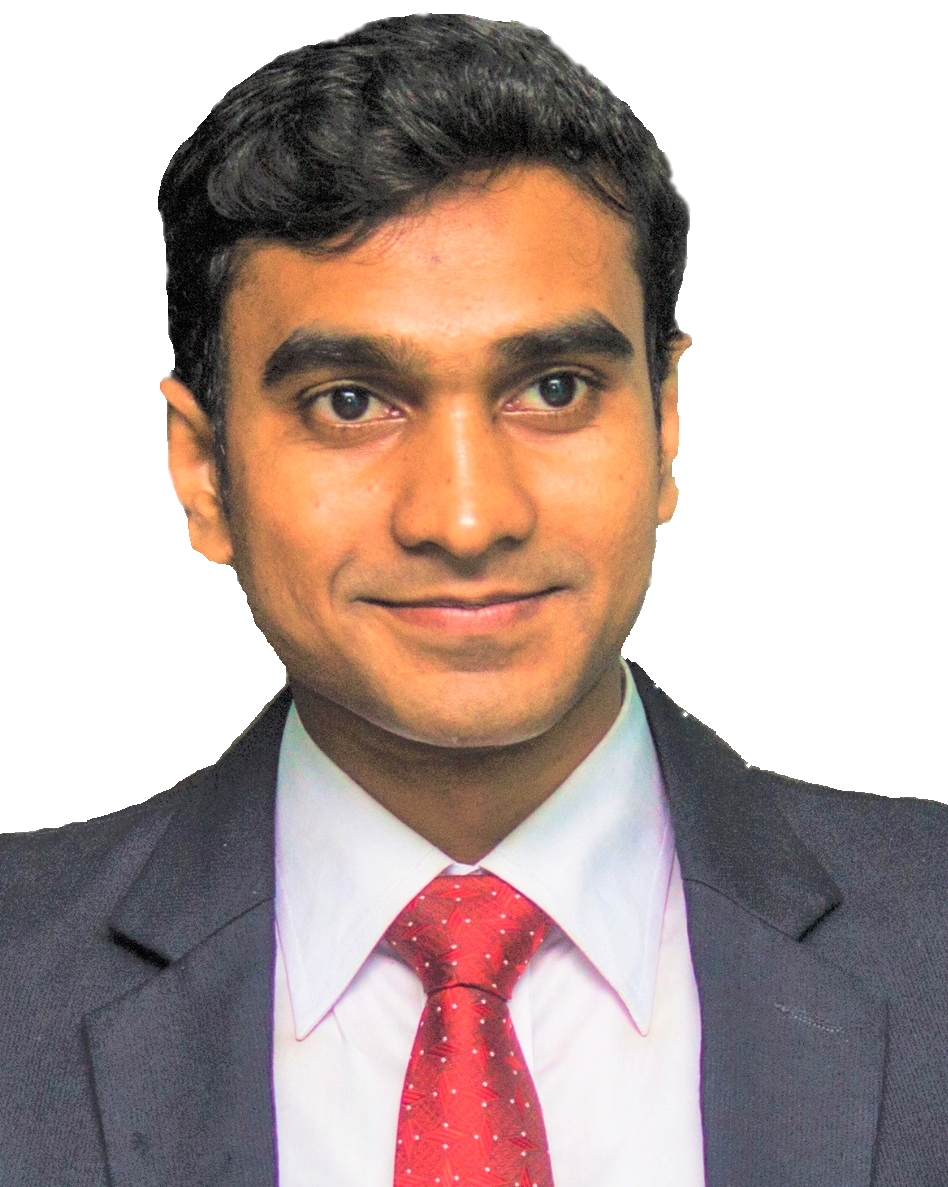}}]{Ziaur Rahman} is PhD candidate in Cyber security of RMIT University, Australia. He served Mawlana Bhashani Science and Technology University, Bangladesh as an Associate Professor in ICT. He casually served RMIT, Monash, Deakin and Charles Sturt University, Australia. Three (03) articles he coauthored were nominated and received the best paper awards. He is affiliated with the IEEE, ACM, Australian Computer Society. His research interests include blockchain technology, security of the internet of things (IoT), intrusion detection, and machine learning.
\end{IEEEbiography}

\begin{IEEEbiography}[{\includegraphics[width=1in,height=1.25in,clip,keepaspectratio]{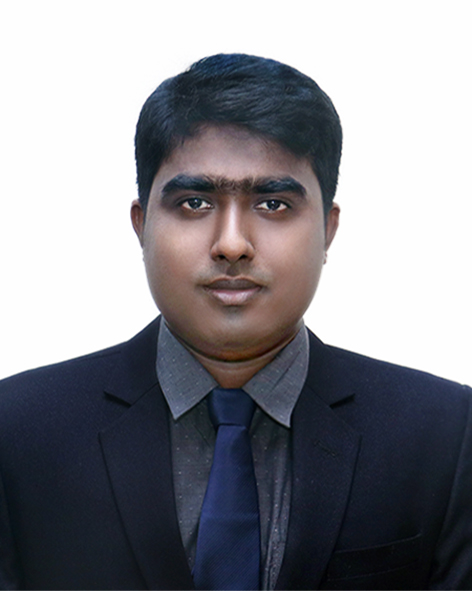}}]{Kawsar Ahmed} received his B.Sc. and M.Sc. Engineering Degree in Information and Communication Technology (ICT) at Mawlana Bhashani Science and Technology University, Tangail, Bangladesh. He has achieved gold medals for engineering faculty first both in B.Sc. (Engg.) and M.Sc. (Engg.) degree from the university for his academic excellence. Currently, he is pursuing his Ph.D. in Electrical Engineering from the University of Saskatchewan, Canada. He is the research coordinator of the "Group of Biophotomati$\chi$".  His research interests include biomedical engineering, sensor design, biophotonics, nanotechnology, data mining, machine learning, and bioinformatics.
\end{IEEEbiography}

\begin{IEEEbiography}[{\includegraphics[width=1in,height=1.25in,clip,keepaspectratio]{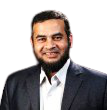}}]{Rafiqul Islam} has been working as an Associate Professor at the School of Computing, Mathematics and Engineering, Charles Sturt University, Australia. Dr Islam’s main research background in cybersecurity focuses on malware analysis and classification, security in the cloud, privacy in social media, and the dark web. Dr. Islam has a strong research background in Cybersecurity with a specific focus on malware analysis and classification, Authentication, security in the cloud, privacy in social media and Internet of Things (IoT). He is leading the Cybersecurity research team and has developed a strong background in leadership, sustainability, collaborative research in the area. He has a strong publication record and has published more than 160 peer-reviewed research papers. His contribution is recognized both nationally and internationally through achieving various rewards such as professional excellence reward, research excellence award, leadership award. Dr. Islam has recognized at the national forefront of his research field ‘cybersecurity’, which is now one of the national/International research priority (financial, political and social aspects). His is Co-Investigator on a successful Cybersecurity CRC  (68 M) to which he is contributing to the projects related to Resilient Networks, Security and configuration management of IoT system, Platform and Architecture of cybersecurity as a service and malware detection and removal. 
\end{IEEEbiography}

\end{document}